\documentclass[structabstract]{aa}  
\usepackage{graphicx}
\usepackage{epsfig}
\usepackage{natbib}
\usepackage{float}
\usepackage{txfonts}
\usepackage[babel=true]{csquotes}
\usepackage{hyperref}
\hypersetup{colorlinks=true,       
    linkcolor=blue,          
    citecolor=blue,        
    urlcolor=blue
    }
\def\PSF{\mathrm{PSF}}
\newcommand{\mynorm}[2][2]{\left\Vert{#2}\right\Vert_{#1}}

\begin{document}
   \title{Evolution of dust in the Orion Bar with \emph{Herschel}\thanks{\emph{Herschel} is an ESA space observatory with science instruments provided by European-led Principal Investigator consortia and with important participation from NASA.}}
   \subtitle{I. Radiative transfer modelling}

   \author{H. Arab\inst{1}, A. Abergel\inst{1}, E. Habart\inst{1}, J. Bernard-Salas\inst{1}, H. Ayasso\inst{1}, K. Dassas\inst{1}, P. G. Martin\inst{2}, \and G. J. White\inst{3,}\inst{4}}

   \institute{Institut d'Astrophysique Spatiale (IAS), UMR8617, CNRS/Universite Paris-Sud, 91405 Orsay, France\\
              \email{heddy.arab@ias.u-psud.fr}
    \and
             Canadian Institute for Theoretical Astrophysics, Toronto, Ontario, M5S 3H8, Canada
    \and
             The Rutherford Appleton Laboratory, Chilton, Didcot OX11 0QX, UK
     \and   
             Department of Physics \& Astronomy, The Open University, Milton Keynes MK7 6 AA, UK
             }

   \date{}

 
  \abstract
   {Interstellar dust is a key element in our understanding of the interstellar medium and star formation. The manner in which dust populations evolve with the excitation and the physical conditions is a first step in the comprehension of the evolution of interstellar dust.}
   {Within the framework of the Evolution of interstellar dust \emph{Herschel} key program, we have acquired PACS and SPIRE spectrophotometric observations of various photodissociation regions, to characterise this evolution. The aim of this paper is to trace the evolution of dust grains in the Orion Bar photodissociation region.}
   {We use \emph{Herschel}/PACS (70 and 160 $\mu$m) and SPIRE (250, 350 and 500 $\mu$m) together with \emph{Spitzer}/IRAC observations to map the spatial distribution of the dust populations across the Bar. Brightness profiles are modelled using the DustEM model coupled with a radiative transfer code.}
   {Thanks to \emph{Herschel}, we are able to probe finely the dust emission of the densest parts of the Orion Bar with a resolution from 5.6$\arcsec$ to 35.1$\arcsec$. 
    These new observations allow us to infer the temperature of the biggest grains at different positions in the Bar, which reveals a gradient from $\sim$ 80 K to 40 K coupled with an increase of the spectral emissivity index from the ionization front to the densest regions. \\
Combining \emph{Spitzer}/IRAC observations, which are sensitive to the dust emission from the surface, with \emph{Herschel} maps, we have been able to measure the Orion Bar emission from 3.6 to 500 $\mu$m. We find a stratification in the different dust components which can be reproduced quantitatively by a simple radiative transfer model without dust evolution (diffuse ISM abundances and optical properties). \\
However including dust evolution is needed to explain the brightness in each band. PAH abundance variations, or a combination of PAH abundance variations with an emissivity enhancement of the biggest grains due to coagulation give good results. Another hypothesis is to consider a different length along the line of sight of the Bar at the ionization front than in the densest parts.}
{}

\authorrunning{H. Arab et al.}
   \maketitle
%

\section{Introduction}

Interstellar grains constantly interact with their surrounding gaseous environment. They also strongly interact with radiation, absorbing the UV/visible photons from stars and re-emitting the energy in the infrared range. Their emission wavelength range depends on both grain size and composition. To study the dust properties and their evolution, many models have been developed in the last two decades in order to reproduce the dust emission spectrum and the extinction curve (e.g., \citealp{1990A&A...237..215D, 2004ApJS..152..211Z, 2007ApJ...657..810D, 2011A&A...525A.103C}). Most of these use three dust components to reproduce the observational constraints: polycyclic aromatic hydrocarbons (PAHs), \enquote{very small grains} (VSGs), and big grains (BGs). The two first are made of carbonaceous particles. Due to their sizes ($\sim [0.4-10]$ nm), they are stochastically heated  and emit most of their energy below $60 \ \mu$m. The BGs (up to $\sim 200$ nm) are in thermal equilibrium with the radiation field and contain most of the dust mass. During their life, dust grains undergo modification of their size, structure and chemical composition. To understand this evolution, we need to trace the characteristics of grains in relation to changes of the physical and dynamical properties. \\
Photo-dissociation regions (PDRs) are the surface layers where the radiation field is able to dissociate $\mathrm{H_{2}}$ molecules but cannot ionize hydrogen atoms. Thus, they are found at the interface of HII regions and molecular clouds. In such zones, the radiation field coming from nearby OB stars regulates the physical and chemical evolution of the gas. That is why PDRs are the favourite laboratories to study evolution of the dust and gas constituents with the local excitation and the physical conditions. \\
\begin{figure*}         
\centering      
\vspace{-2cm}    
\begin{tabular}{c c c}     

\hskip-0.5cm\includegraphics[width=7.2cm]{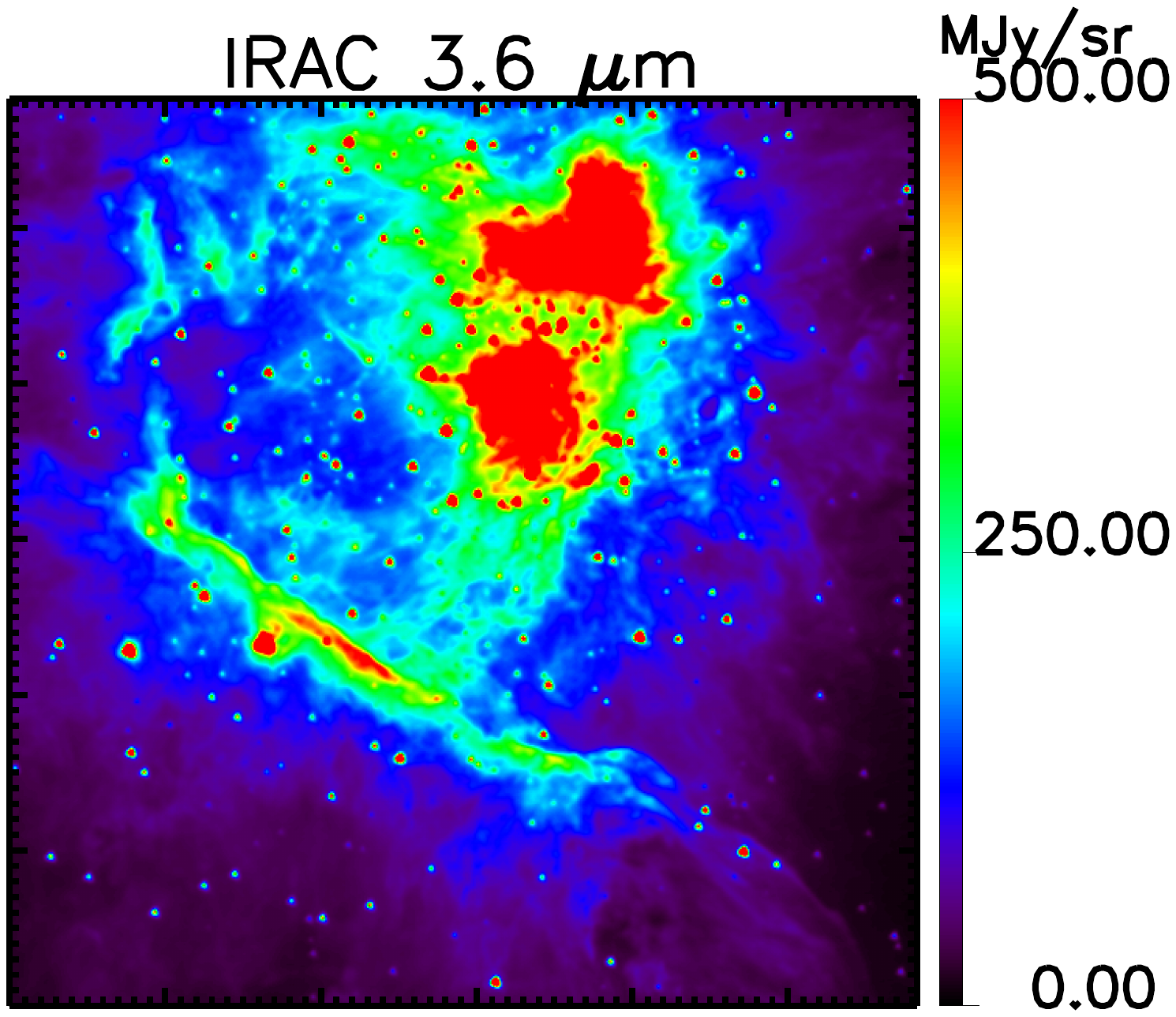} & \hskip-2cm\includegraphics[width=7.2cm]{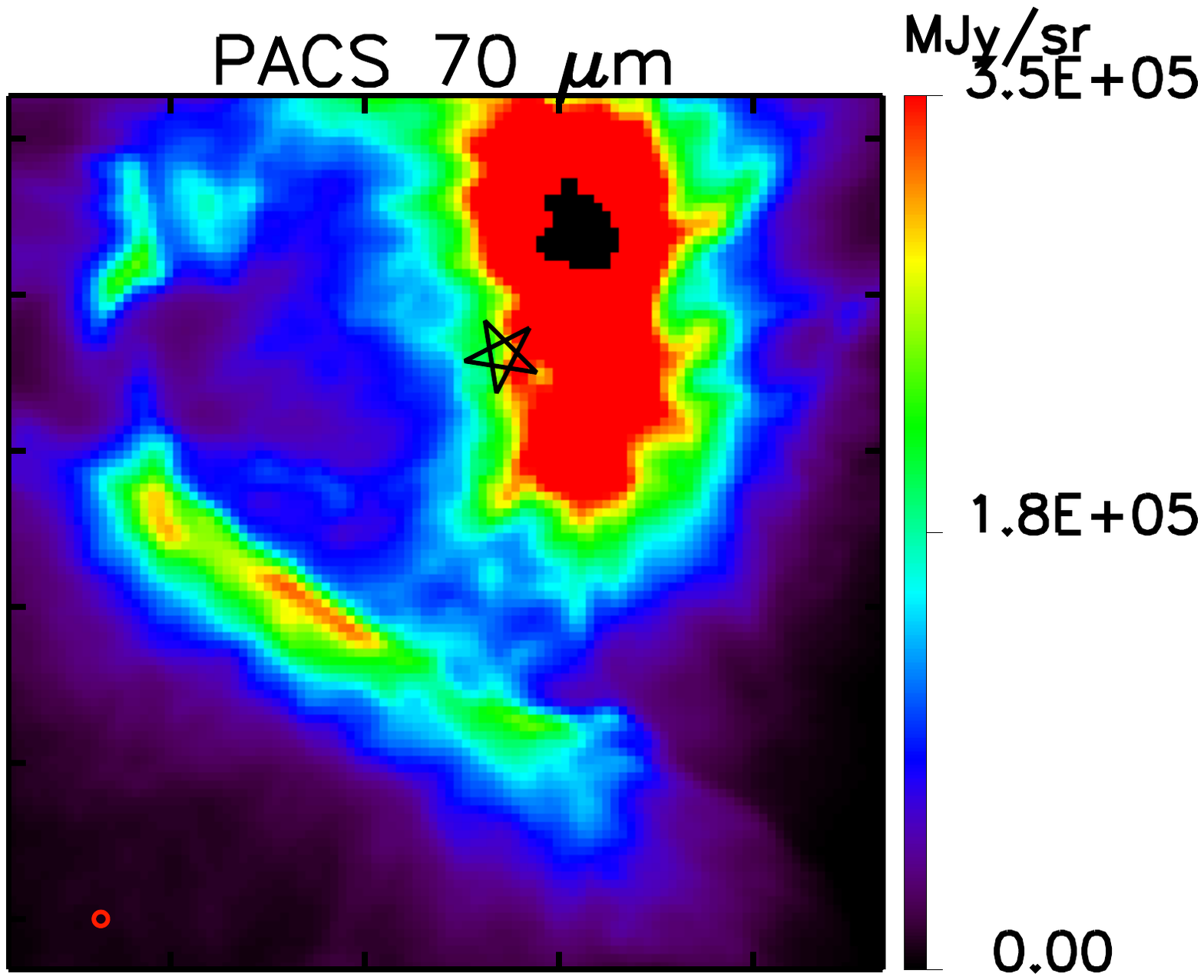} &  \hskip-2cm\includegraphics[width=7.2cm]{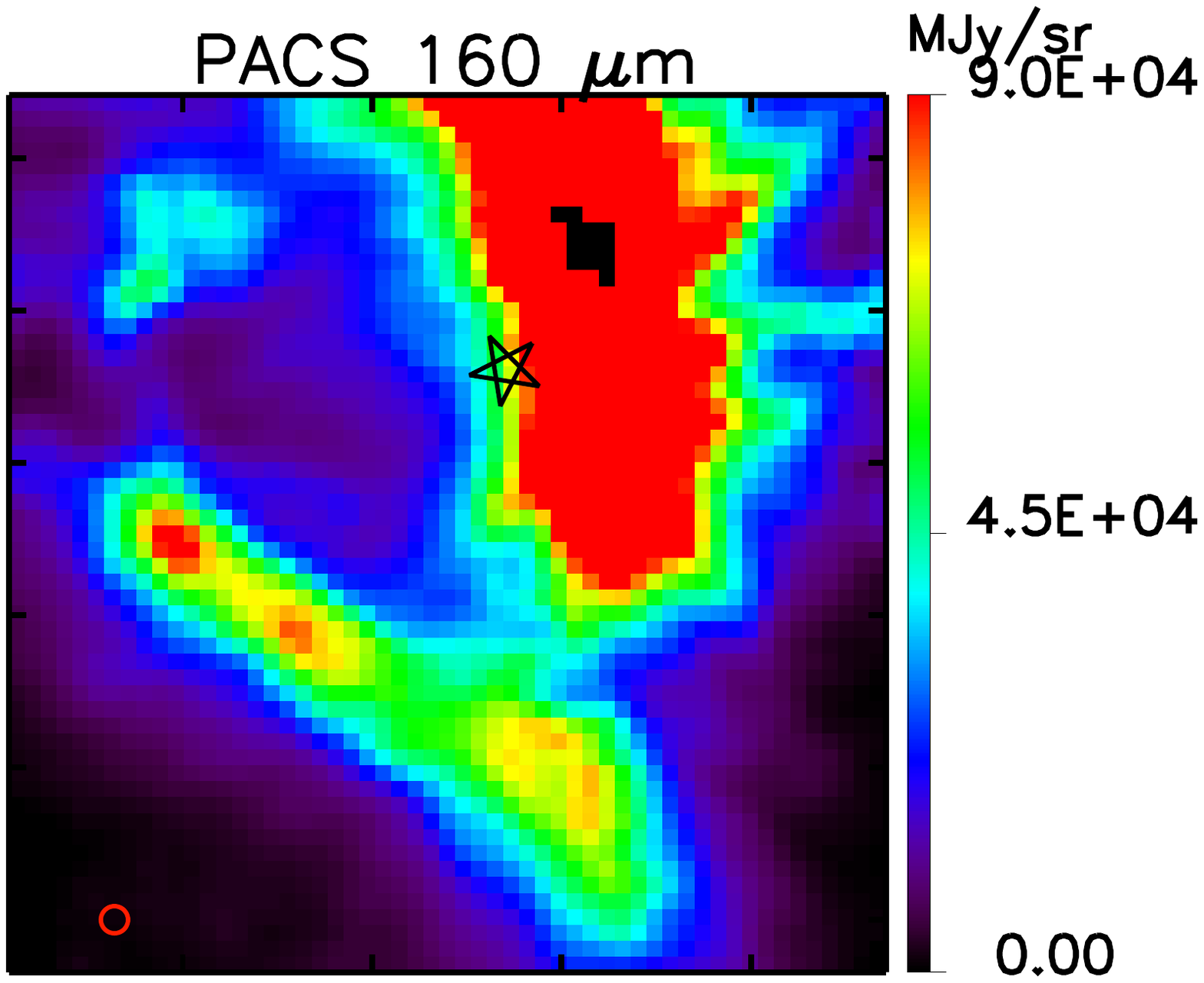} \vspace{-5cm} \\
\hskip-0.5cm\includegraphics[width=7.2cm]{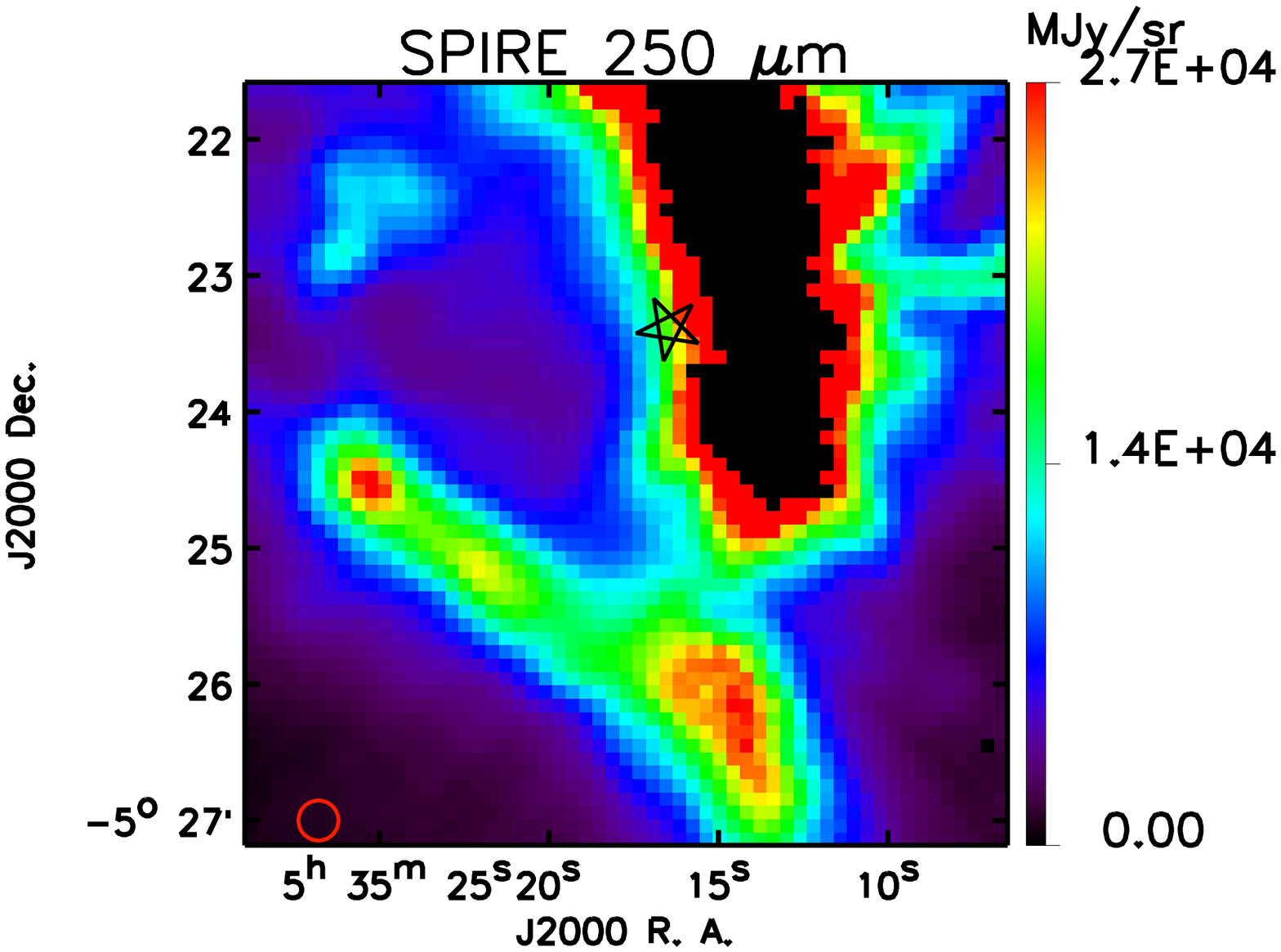} &  \hskip-2cm\includegraphics[width=7.2cm]{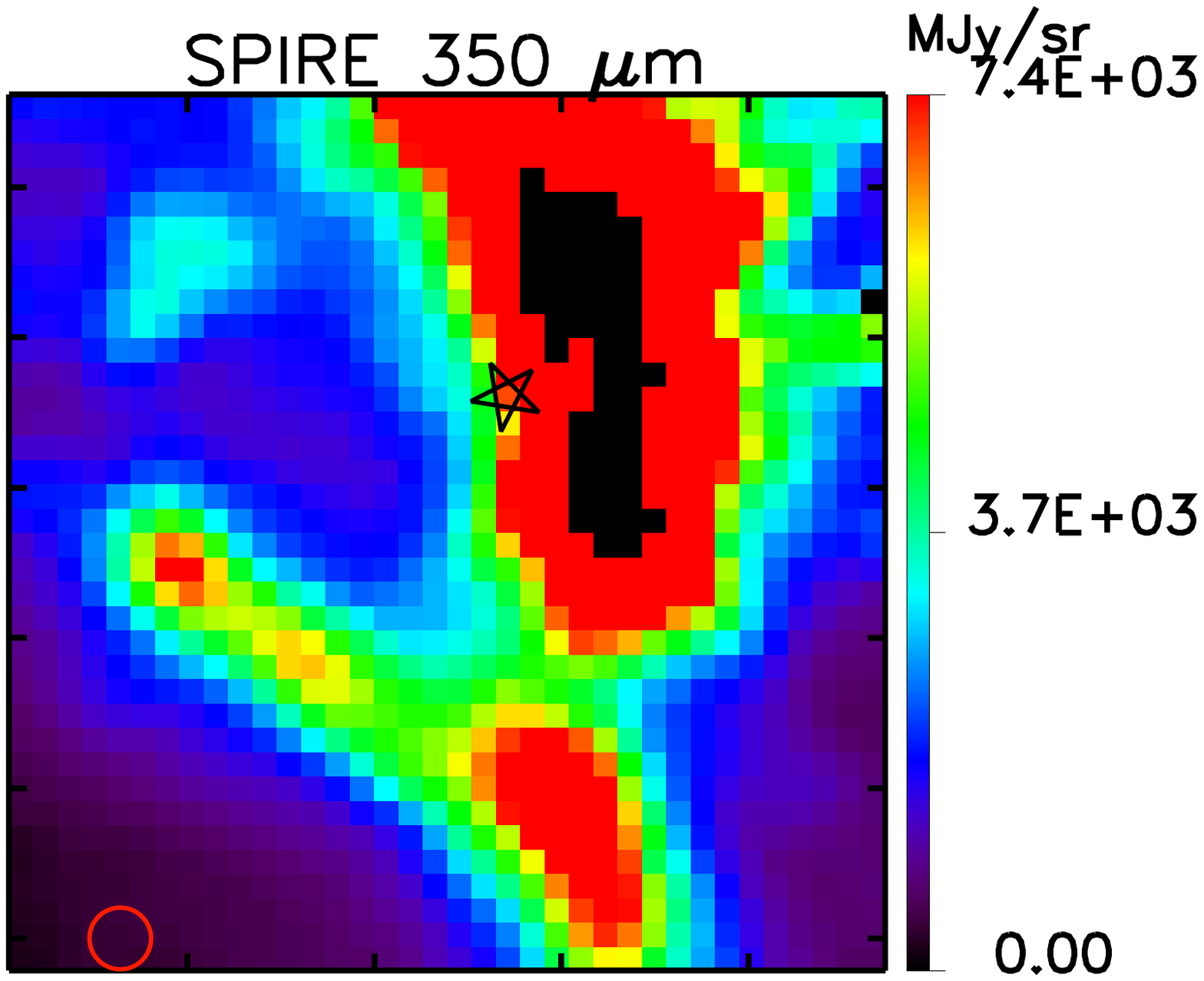} &  \hskip-2cm\includegraphics[width=7.2cm]{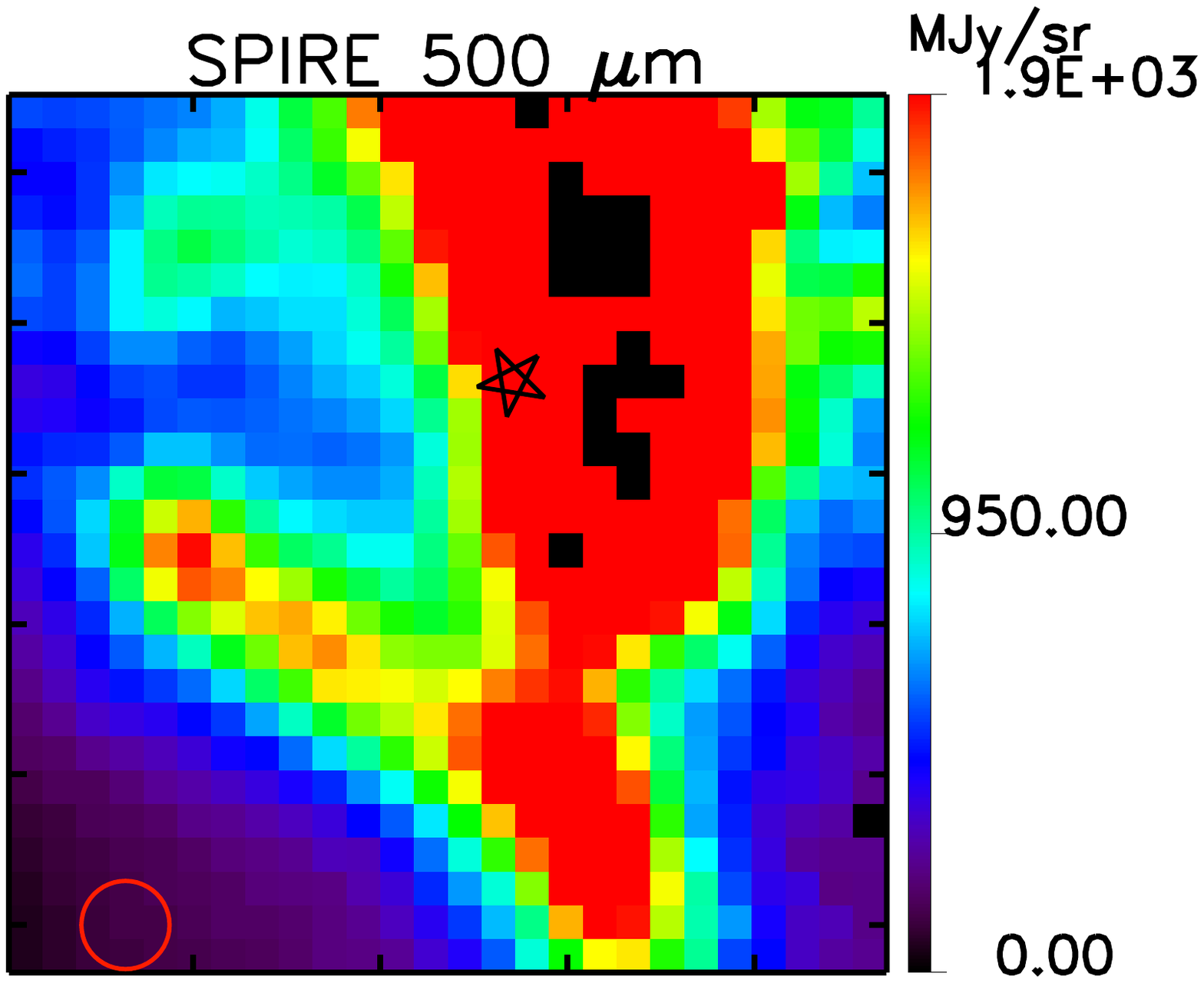}     
\end{tabular}
\vskip-2.5cm
\caption{Orion bar maps observed by IRAC, PACS and SPIRE instruments. The red circle bottom right stands for the FWHM for each channel, and the black star shows the location of the illuminating source. The saturated pixels are indicated in black.}
\label{fig:map}       
\end{figure*}
The Orion Bar has been extensively observed, leading to a large dataset in different tracers of the gas component (e.g., \citealp{1995A&A...299..179W, 1996A&A...313..633V, 2000A&A...364..301W, 2011A&A...530L..16G}). Located between the Trapezium cluster and the Orion Molecular Cloud, the Bar is part of the Orion nebula lying at $414\pm7 \ \mathrm{pc}$ from the Earth \citep{2007A&A...474..515M}.The incident radiation field is dominated by the O6 star $\theta^1$ Ori C and has been estimated to $G=[1-4]\times10^{4} \  G_{0}$ at the ionization front (\citealp{1998A&A...330..696M, 1985ApJ...291..722T}) where $G_{0}$ is the integrated intensity of the standard interstellar radiation field (ISRF) given by \cite{1968BAN....19..421H}. \\
Observational studies of the bright Bar have been accompanied by several modelling studies to determine its physical structure, excitation, and gas emission. Previous studies revealed the presence of different layers within the PDR, with a stratified structure typical of an edge-on geometry (e.g., \citealp{1993Sci...262...86T, 1994ApJ...422..136T, 2003ApJ...597L.145L}). $\mathrm{HCO^{+}}$ and HCN $J=1-0$ emission line maps from \cite{2000ApJ...540..886Y} showed the presence of dense clumps ($n_{H}=3\times10^{6} \ \mathrm{cm}^{-3}$) embedded in an interclump medium with a density $\mathrm{n_{H}}\approx5\times10^{4} \ \mathrm{cm^{-3}}$, which emits most of the intensity.\\

Within the framework of the \enquote{Evolution of interstellar dust} \emph{Herschel} key program \citep{2010A&A...518L..96A}, the Orion Bar has been observed by the ESA \emph{Herschel Space Observatory} \citep{2010A&A...518L...1P}. Its unprecedented spatial resolution together with its spectral coverage at the far-Infrared wavelengths allow us to probe the dust evolution within PDRs. The large sensitivity range and the spectrophotometric capabilities of PACS \citep{2010A&A...518L...2P} and SPIRE \citep{2010A&A...518L...3G} detectors give a unique view of the Orion Bar at the far-Infrared wavelengths. Several papers by our group deal with \emph{Herschel} observations of the Orion Bar. \cite{2012A&A...538A..37B} focus on PACS spectroscopic data whereas Habart et al. (in prep.) explore SPIRE/FTS observations. In this paper, we present the photometric maps from PACS and SPIRE to study the emission of dust at thermal equilibrium with the radiation. We seek to address the following issues: what are the processes involved in the dust evolution in the Orion Bar; are there abundance variations from the diffuse ISM; and are there any differences in the optical properties of the grains in the Bar compared to grains in the diffuse ISM? 

Section \ref{obs} describes the photometric data from SPIRE and PACS, explaining how they have been reduced. In Section \ref{subsect:morph}, the morphology of the Orion Bar in \emph{Spitzer} and \emph{Herschel} bands is presented. These new observations allow us to study both the evolution of the BG spectral energy distribution across the Orion Bar (Sect. \ref{subsect:BGspec}), but also to probe dust emission by mapping its spatial distribution (Sect. \ref{subsect:BP}). In section \ref{modelling}, we model the dust emission in the Orion Bar using the DustEM model \citep{2011A&A...525A.103C} coupled with a radiative transfer code, which is compared to our data. The results are discussed in section \ref{discussion}.  

\newcommand{\lien}[1]{%
   $\,$\footnote{$\,$\url{#1}}}
\section{Observations and data processing}
\label{obs}
The Orion Bar was mapped during the \emph{Herschel} Science Demonstration Phase (SDP) on February 23 and March 11 2010 using the PACS (70 and 160 $\mu$m) and SPIRE photometers (250, 350 and 500 $\mu$m).
For PACS, two concatenated and perpendicular 12'$\times$12' scan maps (corresponding to two obsIds) were observed using the medium scan speed (20"/s), a scan length of 12', a cross-scan step of 50'', and a 15 scan legs (total observing time 1726\,s). For SPIRE, a single map was taken without a repetition (total observing time of 193\,s). 

\subsection{Data processing}

The SPIRE maps reported in this paper are the Level\,2 naive maps delivered by the Herschel Space Center (HIPE version 7.0.1991), with standard corrections for instrumental effects and glitches. Striping induced by offsets in the flux calibration from one detector to another was removed using the Scan Map Destriper module included in the HIPE environment. The overall absolute flux accuracy is dominated by the calibration uncertainty and is conservatively  estimated as $\pm 7\%$\footnote{from the SPIRE observers' manual, available on the webpage: \url{http://Herschel.esac.esa.int/}}. 
\\
\\
The PACS data were processed with HIPE (version 6.0.1196) with special care for deglitching and the removal of the 1/f noise component. For each filter, processing from Level\,0 to Level\,1 is performed on each obsId and yielded one Level\,1 frame by obsId. A second level glitch mask was computed from preliminary maps built from the frames after high-pass median filtering. Object masks were also attached to the Level\,1 frames. These were computed from the preliminary maps (using the second level glitch mask) to flag out the brightest parts of the sky on the high-pass filtering step performed from Level\,1 to Level\,2 (see below).\\
The processing from Level\,1 to Level\,2 combines the two Level\,1 frames in order to obtain one final map for each filter. The 1/f noise component was removed with high-pass median filtering, using the glitch mask and the object mask. In order to avoid any filtering of extended emission, we have taken a relatively broad window size, corresponding to 5 times the scan length. Finally we performed a simple coaddition of the data. The overall absolute flux accuracy is estimated to be $\pm 20\%$ \citep{Ali:2011} for both the blue and red bands corresponding to the conservative value in the case of extended sources.\\
The maps from our \emph{Herschel} programme are shown in Figure\,\ref{fig:map} with the 3.6 $\mu$m IRAC observation (from the Spitzer Data Archive). The zero point of \emph{Herschel} maps is unknown therefore, a background emission centered around the position $\alpha_{J_{2000}}=\mathrm{5h\ 35min\ 26.7s}$ and $\delta_{J_{2000}}=-5^{\circ} 26' 4.7''$ and corresponding to emission unrelated to the Bar was subtracted from the data. This point is not critical for our analysis given the high brightness of the Orion Bar compared to the background emission. The five processed maps are publicly available at \href{http://idoc-herschel.ias.u-psud.fr/sitools/client_user/}{http://idoc-herschel.ias.u-psud.fr/sitools/client\_user/}.

\subsection{Beam effects}
\label{subsect:beams}
\begin{figure}
 \centering
\includegraphics[width=8.5cm]{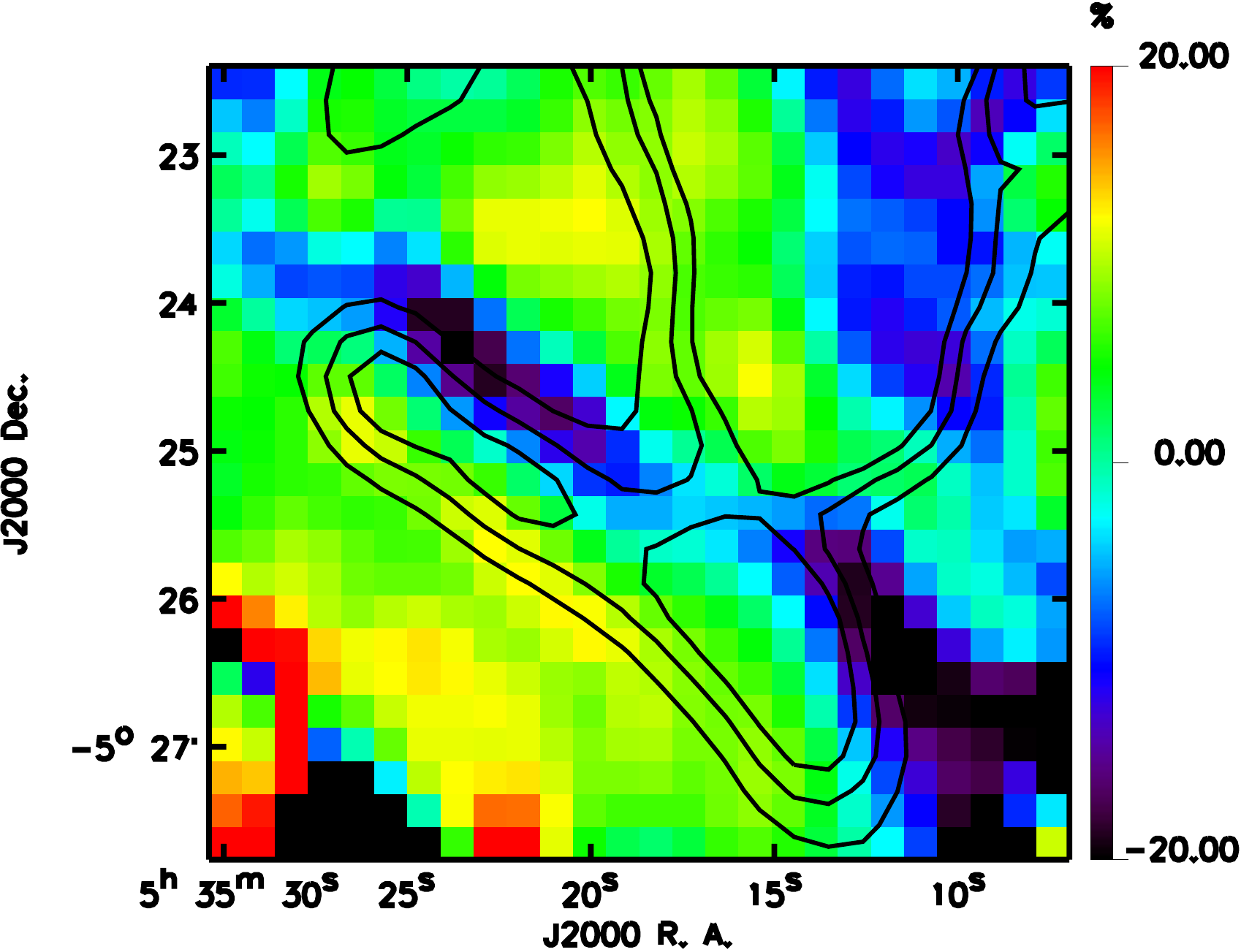}
\caption{$(I_{250\mu m}*h(x,FWHM)-I_{250\mu m}*\hat{k}_{250->500})$: relative differences between the 250\,$\mu$m observation brought to the 500 $\mu$m resolution using a 30.4$\arcsec$ FWHM Gaussian and the transition PSF obtained with our method; black contours show the 250\,$\mu$m emission (levels: 10500, 14000, 17500 MJy.sr$^{-1}$). } 
\label{oribar_beams1}
\end{figure}

To make a coherent spatial study and extract spectral energy distributions (SEDs), all the maps need to be brought in the same beam, in our case the 500\,$\mu$m beam. The simplest way to do that is to assume that \emph{Herschel} beams are Gaussian. Therefore each map is usually convolved by a Gaussian which full width at half maximum (FWHM) is given by:
\begin{equation}
\textrm{FWHM}=\sqrt{\textrm{FWHM}^2_2-\textrm{FWHM}^2_1},
\end{equation}
where FWHM$_1$ is the FWHM of the considered beam and FWHM$_2$ is the FWHM of the larger beam of the dataset.
Nevertheless, simulations and measurements of \emph{Herschel} point spread functions (PSFs) show that the real beams are significantly different from Gaussian since they present asymmetries and secondary lobes \citep{Sibthorpe:2011}. This point is particularly critical since ignoring secondary lobes misestimates the fluxes at the edges of bright structures as illustrated in Figure\,\ref{oribar_beams1}. Thus, it is necessary to take into account the real shape of the beam in our study and we use the theoretical PSFs delivered by the \emph{Herschel} Science Centre. The transition functions needed to convolve the maps to a common beam size does not have an analytical form, and must be numerically calculated. 

\begin{figure}[!ht]
\centering
\includegraphics[width=8.5cm]{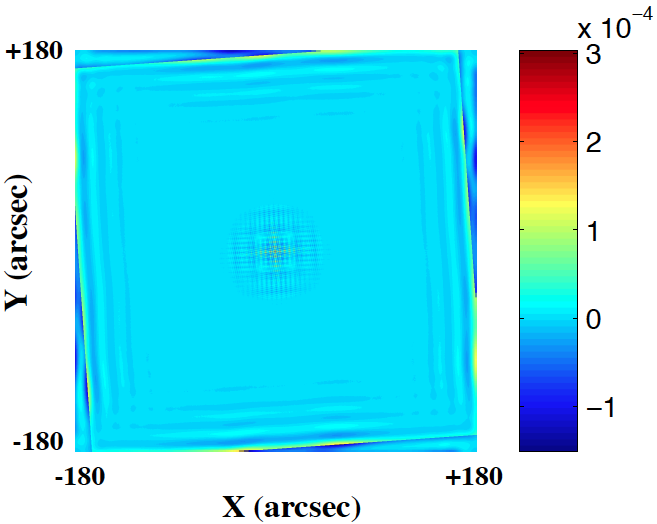}
\caption{Relative error in the reconstruction of the SPIRE 500 $\mu$m from the 250\,$\mu$m one using our inversion method: $(\mathrm{PSF_{500}-PSF_{250}*\hat{k}_{250->500}})$/max($\mathrm{PSF}_{500})$.} 
\label{oribar_beams2}
\end{figure}

\begin{figure*}[!ht]         
\centering      
\vspace{-2cm}    
\begin{tabular}{c c c}     
\hskip-0.5cm\includegraphics[width=7.2cm]{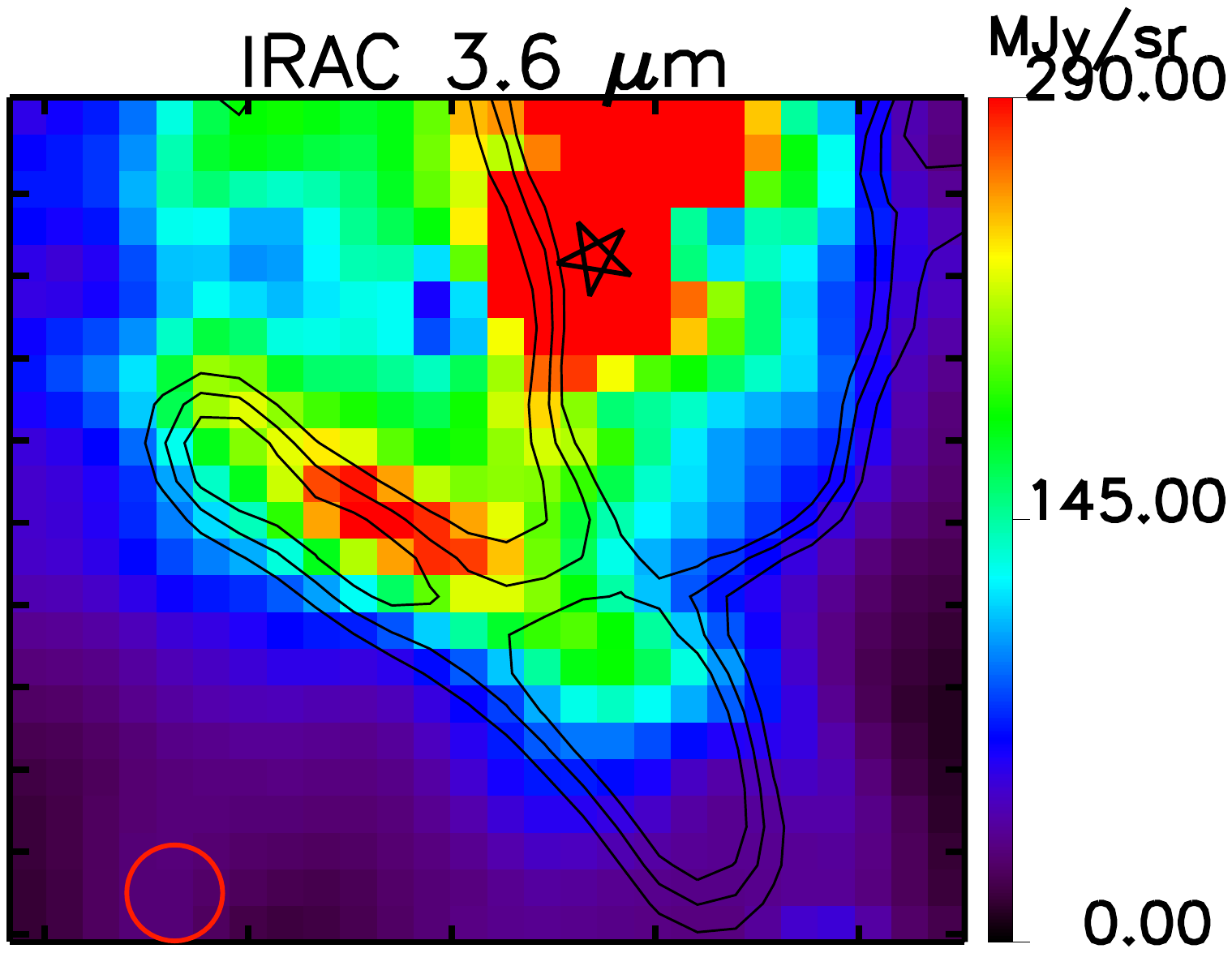} & \hskip-2cm\includegraphics[width=7.2cm]{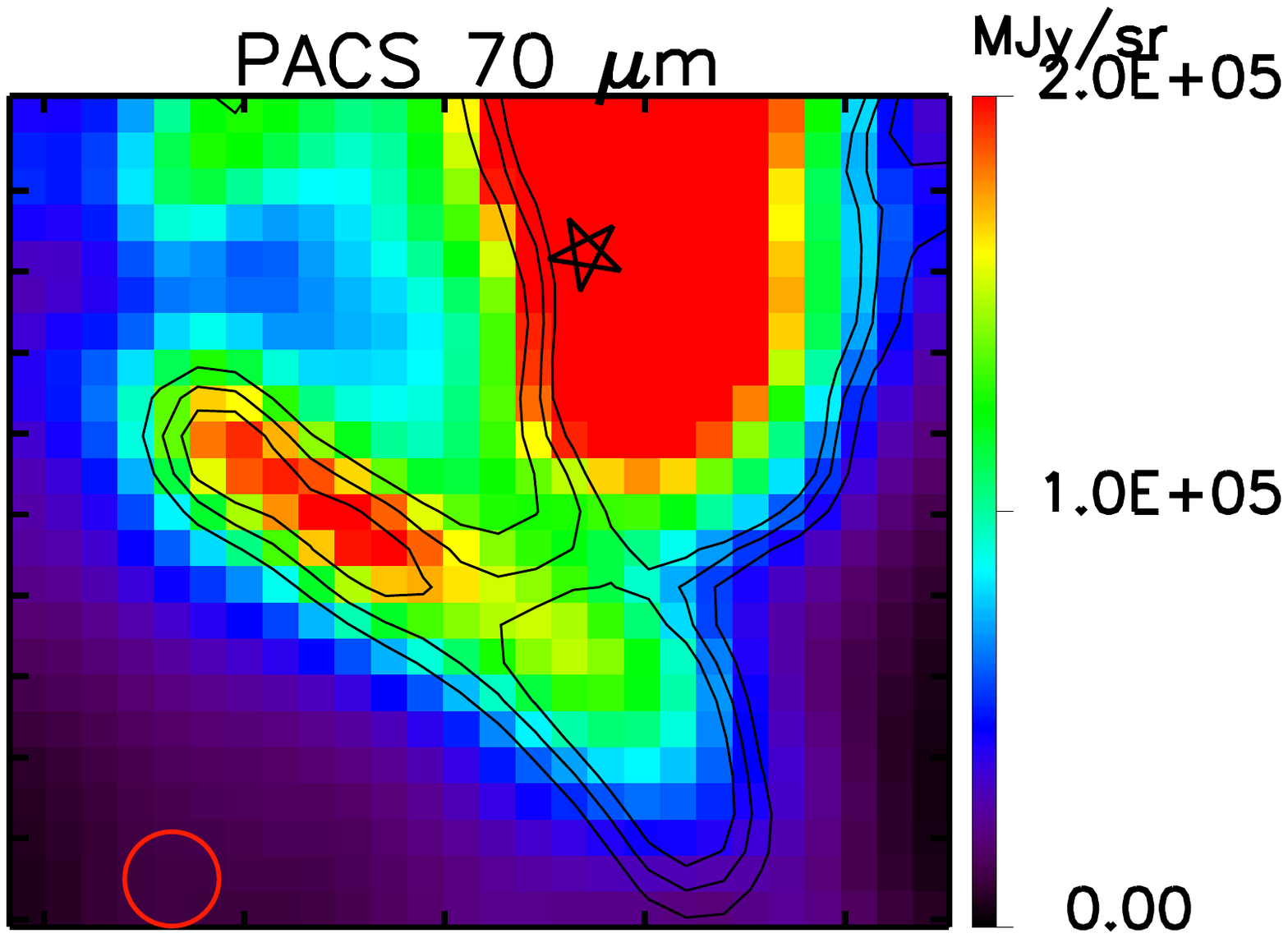} &  \hskip-2cm\includegraphics[width=7.2cm]{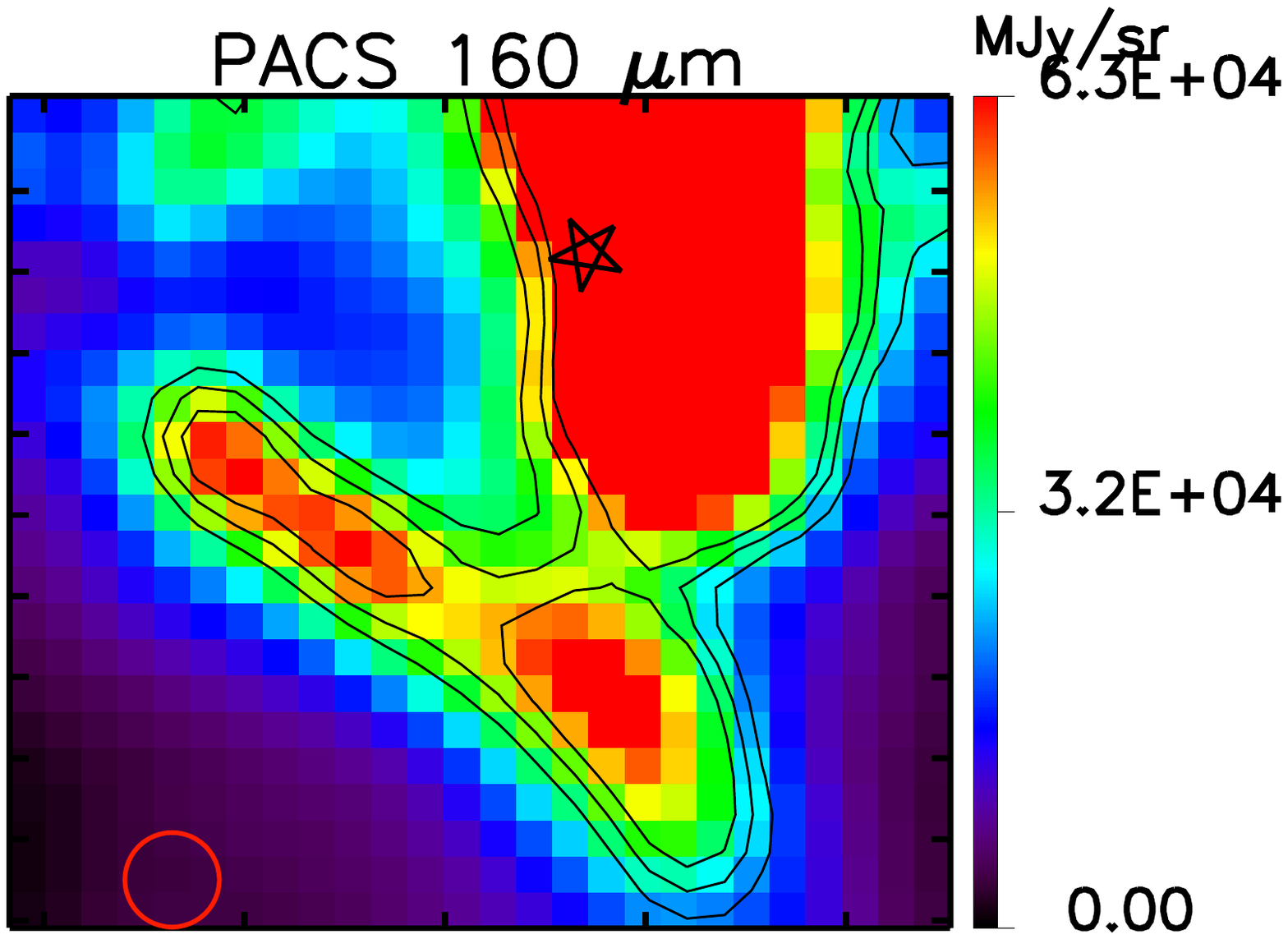} \vspace{-5cm} \\
\hskip-0.5cm\includegraphics[width=7.2cm]{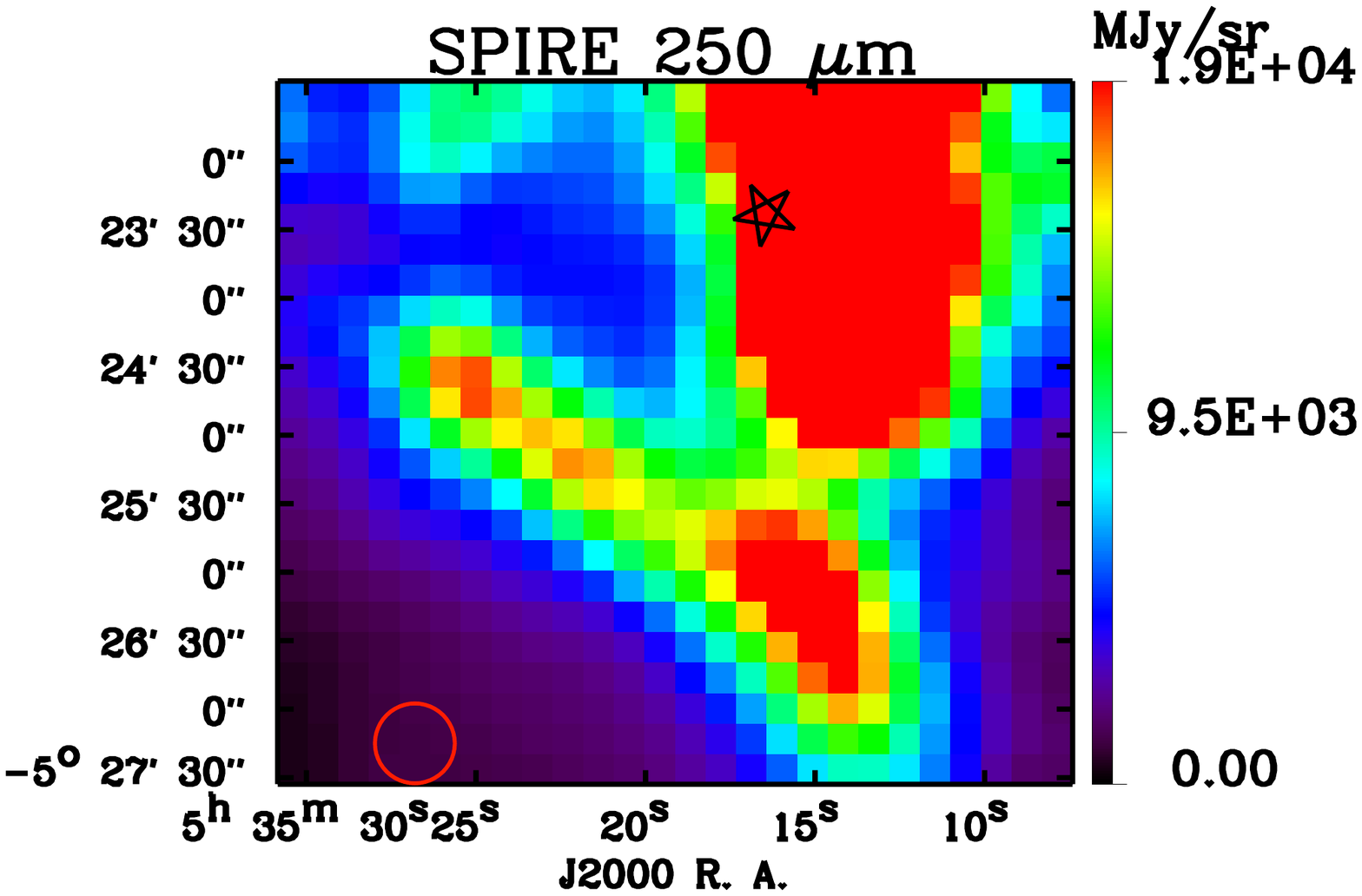} &  \hskip-2cm\includegraphics[width=7.2cm]{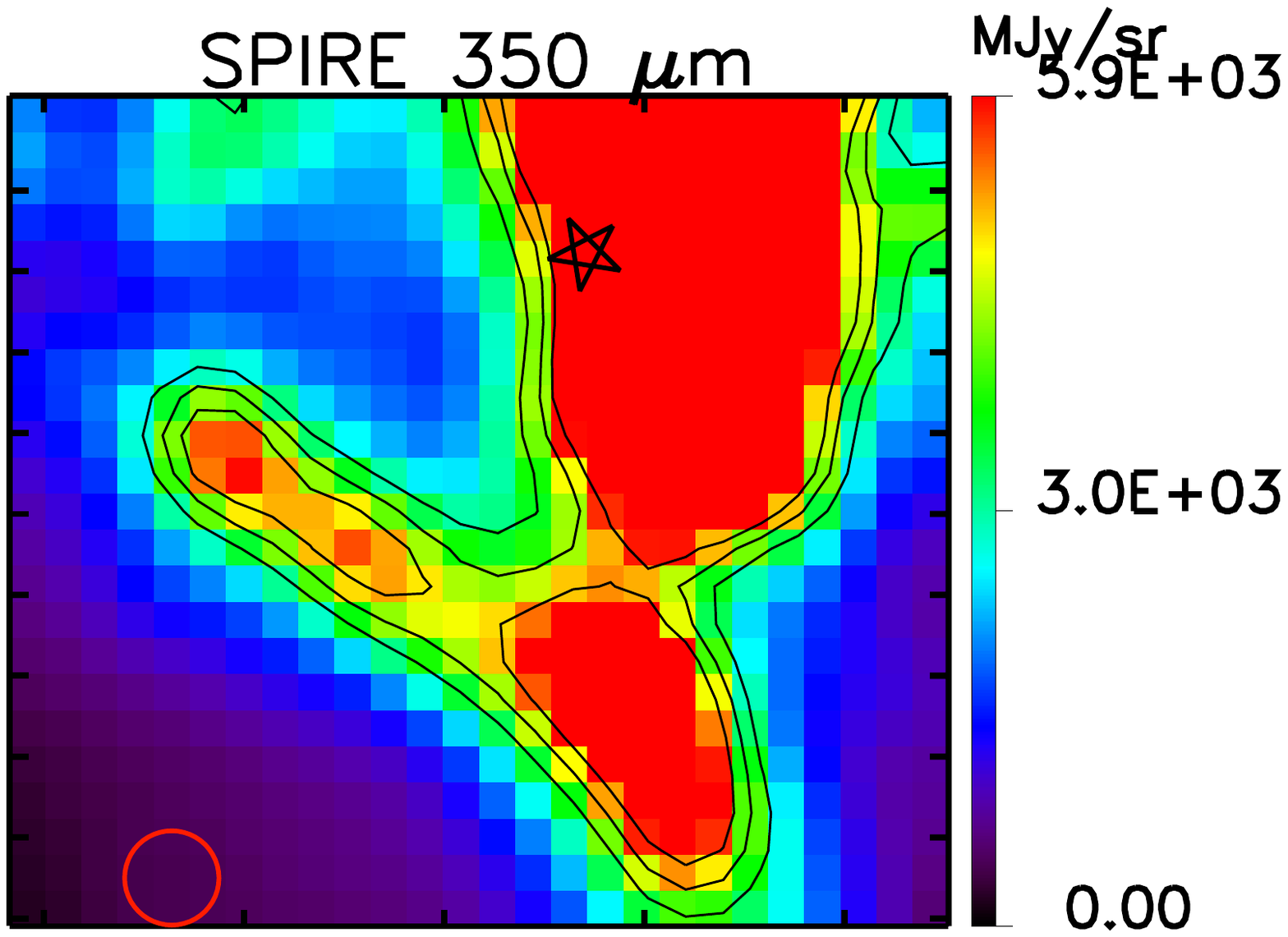} &  \hskip-2cm\includegraphics[width=7.2cm]{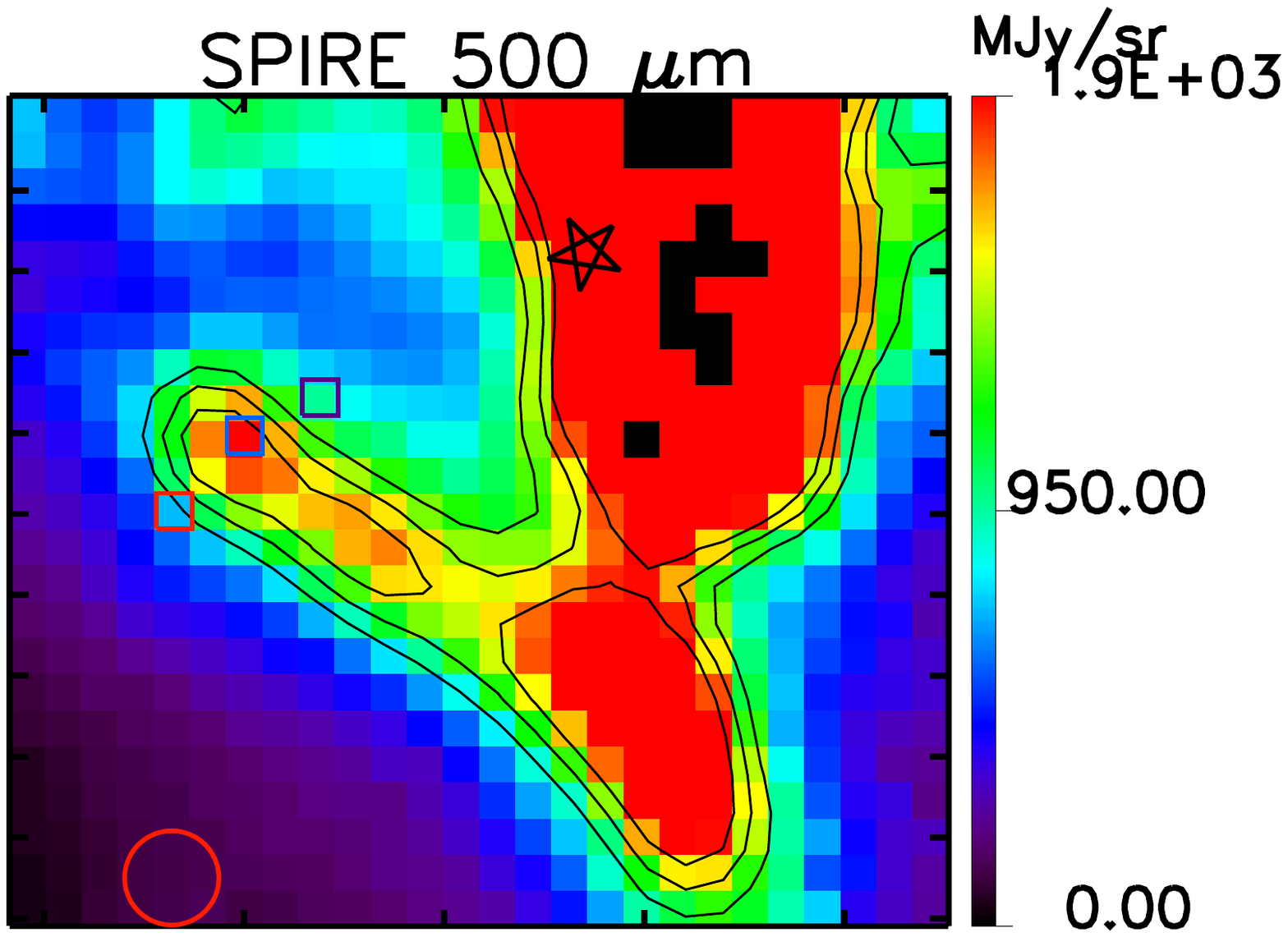}     

\end{tabular}
\vskip-2.5cm
\caption{Observed maps convolved to the 500\,$\mu$m beam (red circle; FWHM=35.1") using the inversion method described section \ref{subsect:beams}. The three colored squares on the 500\,$\mu$m observation stand for the positions of the three SEDs shown Figure \ref{Fig:bgspec}. Black contours show the 250\,$\mu$m emission convolved to the 500\,$\mu$m beam (red circle) with levels at 10000, 12500 and 15000 $\mathrm{MJy.sr^{-1}}$. }
\label{fig:mapconv}       
\end{figure*}

Computing a transition function $k $ between two PSFs, $\PSF_2= \PSF_1*k$, is a deconvolution problem \citep{gonzalez1992}. We choose a regularized least-square method \citep{tikhonov1963} to compute a transition PSF $k$ from a given channel of ${\PSF_{1}}$ to ${ \PSF_{2}}$:
\begin{equation}
\hat{k}= {\arg\min} J(k)=\mynorm{\PSF_{2} - \PSF_{1}*k}^2+\mu \mynorm{\Omega(k)}^2,
\label{eq:PSF}
\end{equation}
where $\Omega$ is a smoothness function taken as a derivative in our case, and $ \mu$ is the regularization parameter which can be fixed according to the noise level present in both PSFs. A transition PSF corresponds to a couple of channels for a given observation because the orientation angles of the instruments are required in the computation.
Recently, \cite{2011PASP..123.1218A} have proposed common-resolution convolution kernels based on another method. In their approach, the low pass filtering, needed to avoid problems in the high frequency component, is made by a filter which shape is fixed whatever the considered beams. In our method the filtering is optimized by adapting the filter according to the shape of the considered PSFs, thus the transition function computed is optimal in the least-squares sense.
Thanks to regularization process, this method is able to give a very good reconstruction of $k$, respecting the balance between spectral information available in the PSFs and the noise levels (Fig. \ref{oribar_beams2}). We present the \emph{Herschel} observations convolved to the 500 $\mu$m resolution and gridding in Figure\,\ref{fig:mapconv}. 

\section{Morphology of the Orion Bar with \emph{Herschel}}
\label{subsect:morph}
We have mapped the emission of the BGs, which trace the matter up to densest regions, of the Orion Bar. Indeed, with \emph{Spitzer}/IRAC, we only detect the emission of the smallest dust particles which is strongly dependent on the radiation field since they are stochastically heated. Therefore, we observe the emission coming from the surface of dense regions. This is illustrated by the first panel of Figure \ref{fig:map}. The 3.6\,$\mu$m map from \emph{Spitzer}/IRAC exhibits numerous narrow bright filaments due to the illuminated edge of the Bar. These bright filaments contain small scale structures, revealing the complexity of the edge of the Bar. \\
\\
In \emph{Herschel} bands, it is striking to note that the global structure of the Bar appears the same (Fig. \ref{fig:map}), especially if we look at the maps brought to the same resolution (Fig. \ref{fig:mapconv}). Even if the ridge appears broader because of resolution effects, the shape is the same and we recognize the sub-structure along the Bar seen with IRAC. 
 However, as we are sensitive to BGs and denser matter, we see several differences in the sub-structure. For example, the South-West part of the Bar ends with a bright cloud invisible at shorter wavelengths, indicating the presence of cold material. Besides, the sub-structures detected along the Bar have not the same relative brightness at different wavelengths. The brightest one at 70 $\mu$m is the faintest at 250 $\mu$m, revealing a strong difference of temperature.
Finally, we observe with \emph{Herschel} an increase of the distance between the observed Bar and the illuminating star with increasing wavelength.\\ 
\\
With these new \emph{Herschel} observations, combined with \emph{Spitzer}/IRAC data, we are able to study the spectrum of the BG emission and to make a coherent spatial study of the different dust populations in the Orion Bar.
 \begin{table}[ht]
      \caption[]{Contribution of VSGs to the overall intensity (in $\%$) according to the DustEM model\tablefootmark{a} in the five \emph{Herschel} bands for the spectra shown in Fig.\ref{Fig:bgspec}.}
         \label{tab:contrib}
              	\vskip-0.5cm
	\renewcommand{\arraystretch}{1.25}
     $$
         \begin{array}{|p{0.49\linewidth}|c|c|c|c|c|}
            \hline
            Offset from the 500\,$\mu$m peak position (5\,h\,35\,min\,16\,s;-5$^{\circ}$23$\arcmin$23$\arcsec$) &  70 \mu m   & 160 \mu m   & 250 \mu m   & 350 \mu m   & 500 \mu m \\
            \hline
             (+28\arcsec,-14\arcsec) (top panel of Fig\,\ref{Fig:bgspec}) & 4.1   & 3.5 & 3.6 & 4.0 & 5.4 \\
             \hline
            (0\arcsec,0\arcsec)  (middle panel of Fig\,\ref{Fig:bgspec})   & 4.0   & 2.4 & 2.3 & 2.6 & 2.7 \\
            \hline
            (-28\arcsec,+28\arcsec) (low panel of Fig\,\ref{Fig:bgspec})   & 4.2   & 2.0 & 1.9 & 2.1 & 2.2  \\
            \hline
         \end{array}
     $$ 
     \vspace{-0.5cm}
     \tablefoot{
     \tablefoottext{a}{We use the same dust populations as in \cite{2011A&A...525A.103C} and an excitation corresponding to $\theta^1$ Ori C.}
     }
   \end{table}

\section{BG spectrum}
\label{subsect:BGspec}

The PACS and SPIRE data allow us to derive the BG component spectrum at each position on the Bar.  After convolving all maps to the 500\,$\mu$m beam, we fit a modified blackbody to the spectra extracted according to:
\begin{equation} 
I_{\nu}=\tau_{\nu_{0}}\left(\frac{\nu}{\nu_{0}}\right)^{\beta}B_{\nu}(T),
\end{equation}
where $I_{\nu}$ is the specific intensity, $\tau_{\nu_{0}}$ is the dust optical depth at frequency $\nu_{0}$, $\beta$ is the spectral emissivity index, $B_{\nu}$ is the Planck's function, and $T$ is the dust temperature.
The fits are performed using the \emph{MPFIT} IDL function \citep{2009ASPC..411..251M}, which relies on the Levenberg-Marquardt algorithm. The free parameters $\tau_{\nu_{0}}$, $\beta$ and $T$ are computed for each pixel of the maps. The five bands are included in the fits since BGs dominate in all bands. This has been verified with DustEM, which is able to quantify the contribution of non-equilibrium emission at a given wavelength\,(Tab.\ref{tab:contrib}). 
 
\begin{figure}[!ht]
   \centering
   \includegraphics[width=\columnwidth]{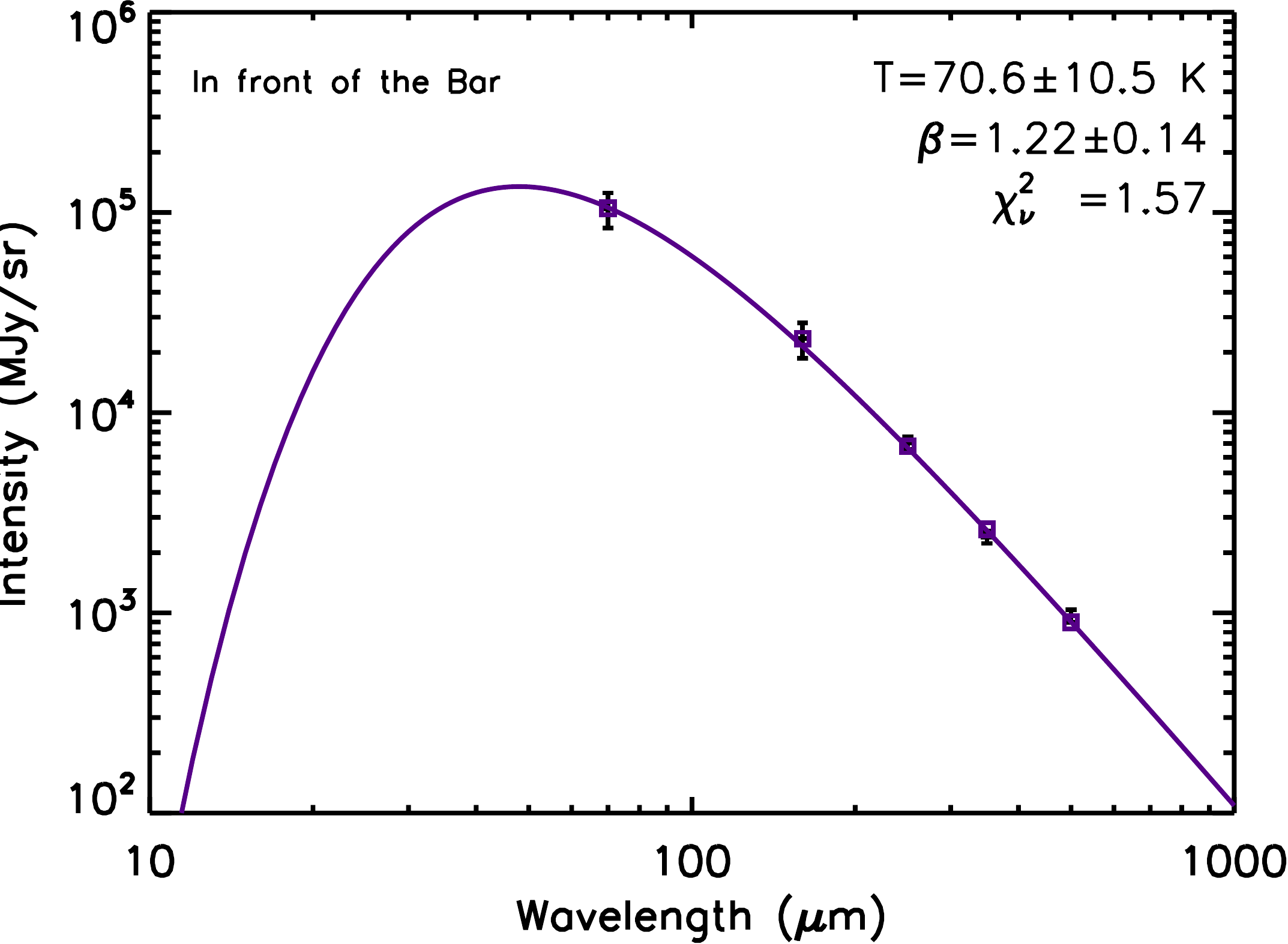}\\
   \includegraphics[width=\columnwidth]{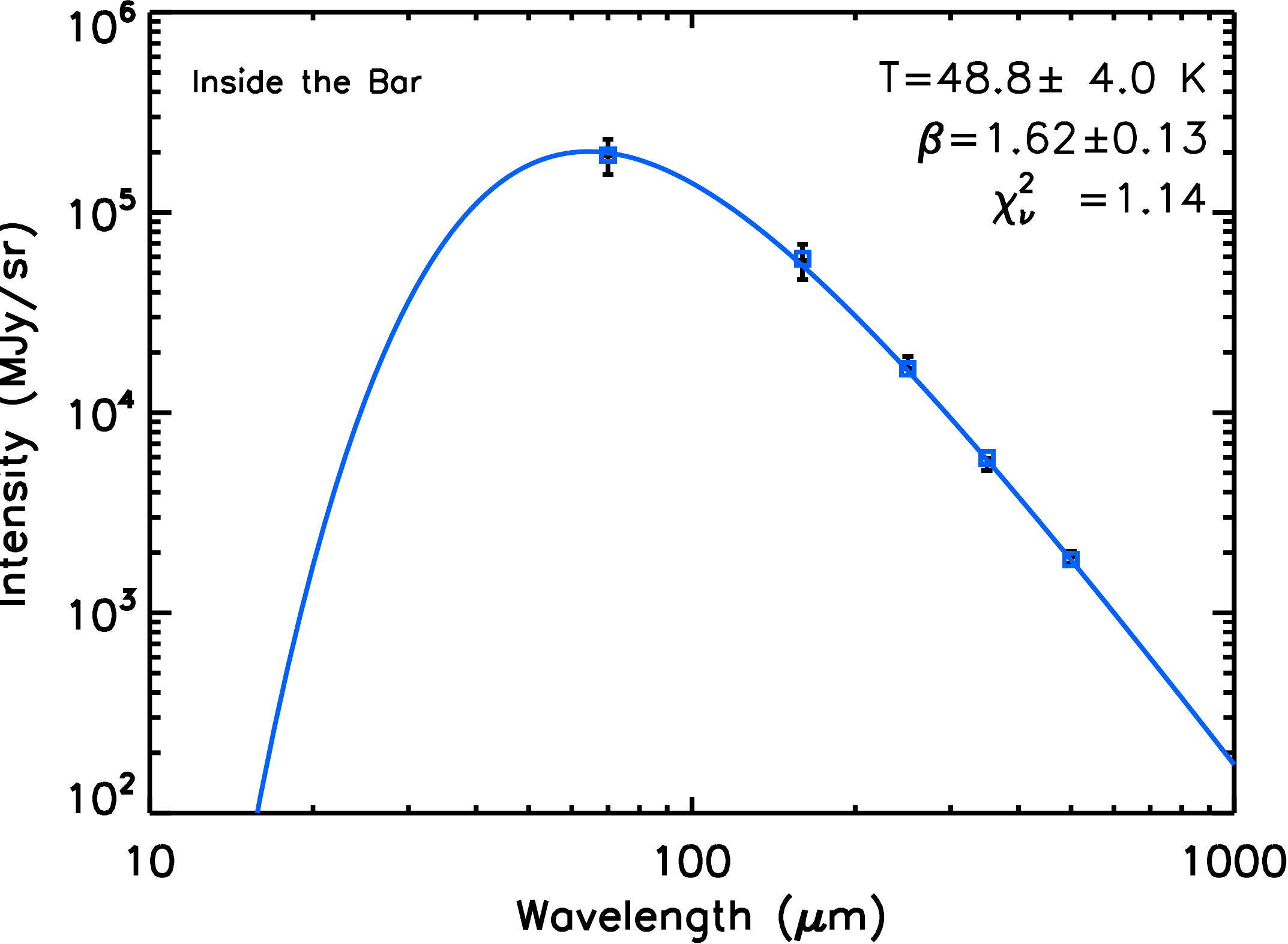}
   \includegraphics[width=\columnwidth]{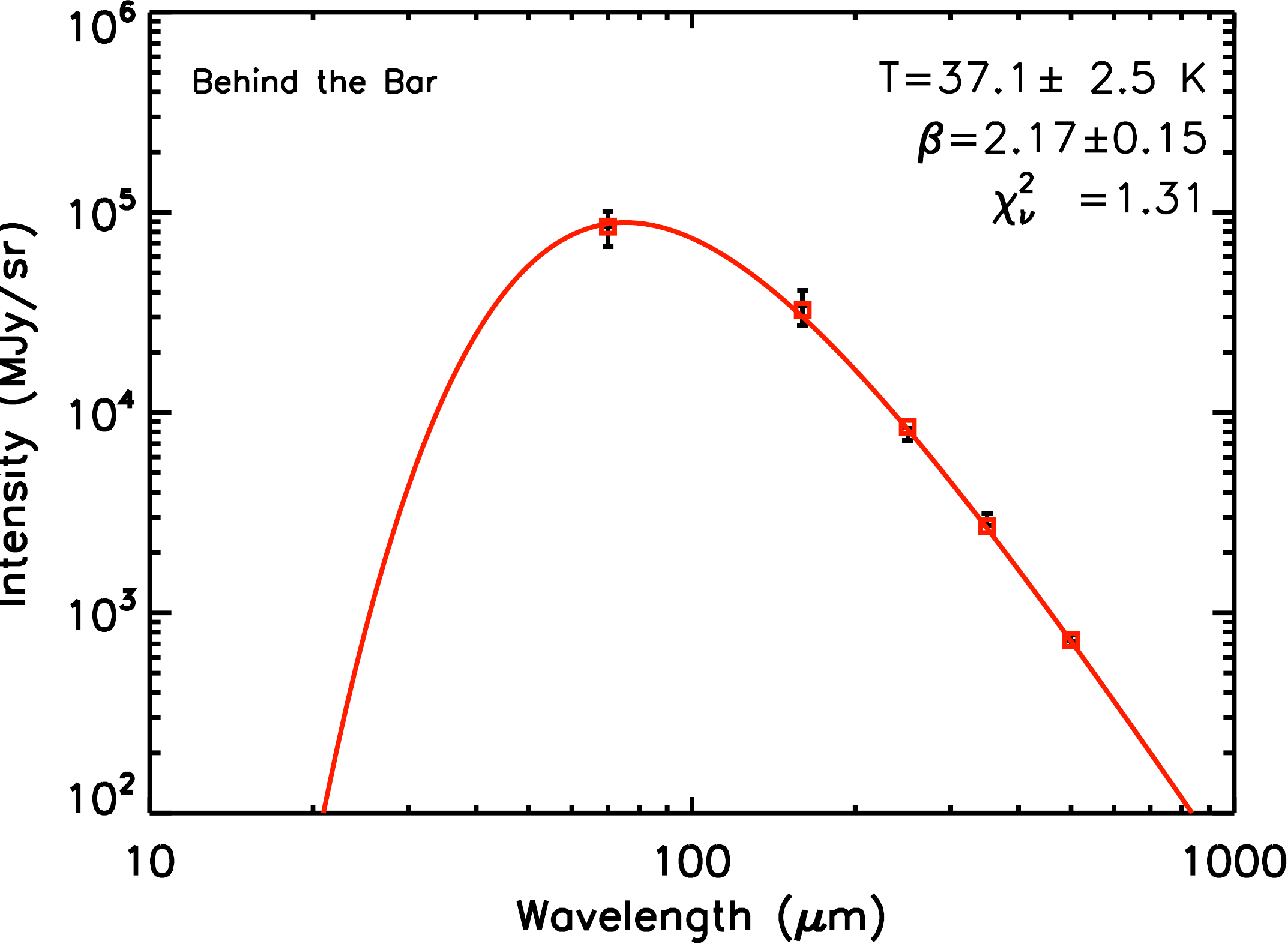}
     \caption{Modified blackbodies fits at different positions across the Bar. Reduced $\chi^2$ is indicated on each fit. The position of each SED is indicated in the 500\,$\mu$m panel of Figure \ref{fig:mapconv}.} 
         \label{Fig:bgspec}
   \end{figure}

Errors entered in the fits are conservative and correspond to the calibration uncertainties (20$\%$ for PACS and 7$\%$ for SPIRE) since statistical noise is negligible compared to calibration errors. The fitted spectra are integrated over the instrument filters so as to take the colour correction into account. We can then study the behaviour of the BG spectrum at different positions in the cloud. Figure \ref{Fig:bgspec} shows the evolution of the BG spectrum while crossing the PDR. The exact positions are shown by the colored squares in Figure\,\ref{fig:mapconv}. In front of the bar (top panel of Fig.\,\ref{Fig:bgspec}), the fitted temperature is $81\pm25$ K whereas as we enter into the bright ridge, it decreases sharply to 47.5$\pm$ 7 K (middle panel of Fig.\,\ref{Fig:bgspec}). 
Behind the Bar (low panel of Fig.\,\ref{Fig:bgspec}), BGs cool down to 39$\pm$4 K. The $\beta$ spectral emissivity index follows an inverse behaviour, increasing from 1.10 to 2.08. This trend is real but might be amplified by the mixture of grains of different temperature along the line of sight. Indeed, mixing can broaden and flatten significantly the modified blackbody shape. This effect is more important in front of the Bar, where the geometry is not edge-on (purple spectrum). Thanks to the high resolution of \emph{Herschel} observations, we are able to probe $T$ and $\beta$ variations on small spatial scales ($\sim$35$\arcsec$), in a single and transition object for the first time. A full study of $T$ and $\beta$ in the Orion Bar region will be presented in detail and compared to previous results in a second paper (Arab et al. in prep).\\
 
\section{Brightness profiles}
\label{subsect:BP}
Using the \emph{Herschel} data coupled with \emph{Spitzer}/IRAC observations, we have been able to study the spatial variations of the different dust population. The BG emission is probed by \emph{Herschel} bands whereas the 3.6 $\mu$m emission from IRAC is used as a proxy of the PAHs. The 8\,$\mu$m channel, usually taken as a tracer of aromatic emission, is affected by saturation. Besides, \emph{Herschel} resolution offers the opportunity to resolve the BG emission within the Orion Bar.

   \begin{figure}
   \centering
   \includegraphics[width=\columnwidth]{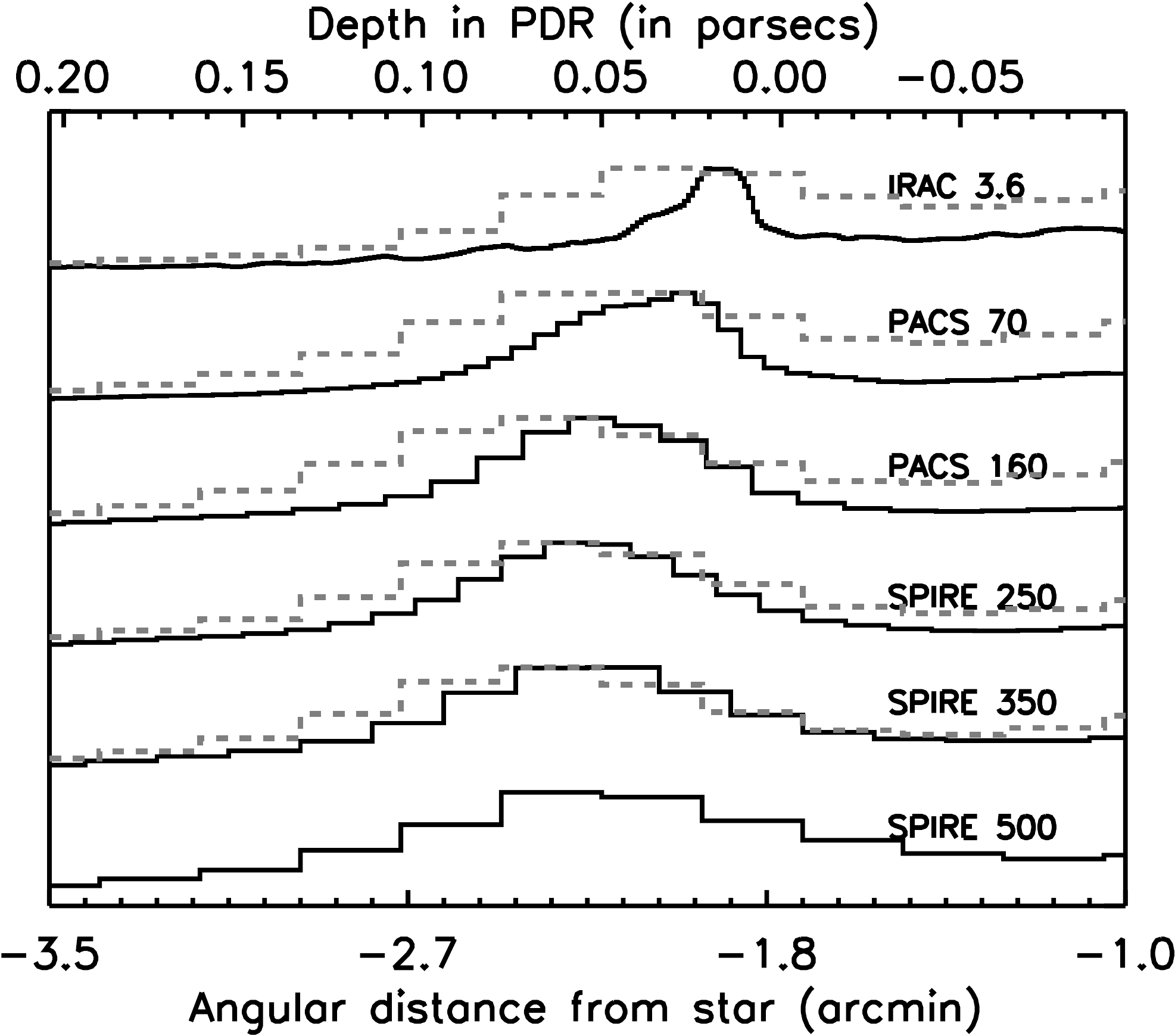}
      \caption{Normalized brightness profiles across the Orion Bar at native resolution (black solid lines) and convolved to the 500\,$\mu$m (grey dashed lines). The resolution of the solid line profiles corresponds to the nominal resolution of each instrument: 1.88$\arcsec$ for IRAC, 5.6$\arcsec$ for PACS 70 $\mu$m, 11.3.$\arcsec$ for PACS 160 $\mu$m, 17.6$\arcsec$ for SPIRE 250\, $\mu$m, 23.9$\arcsec$ for SPIRE 350 $\mu$m, 35.1$\arcsec$ for SPIRE 500\,$\mu$m.}
         \label{Fig:profile}
   \end{figure}

\begin{figure}
 \centering
 \vskip+0.5cm
   \includegraphics[width=\columnwidth]{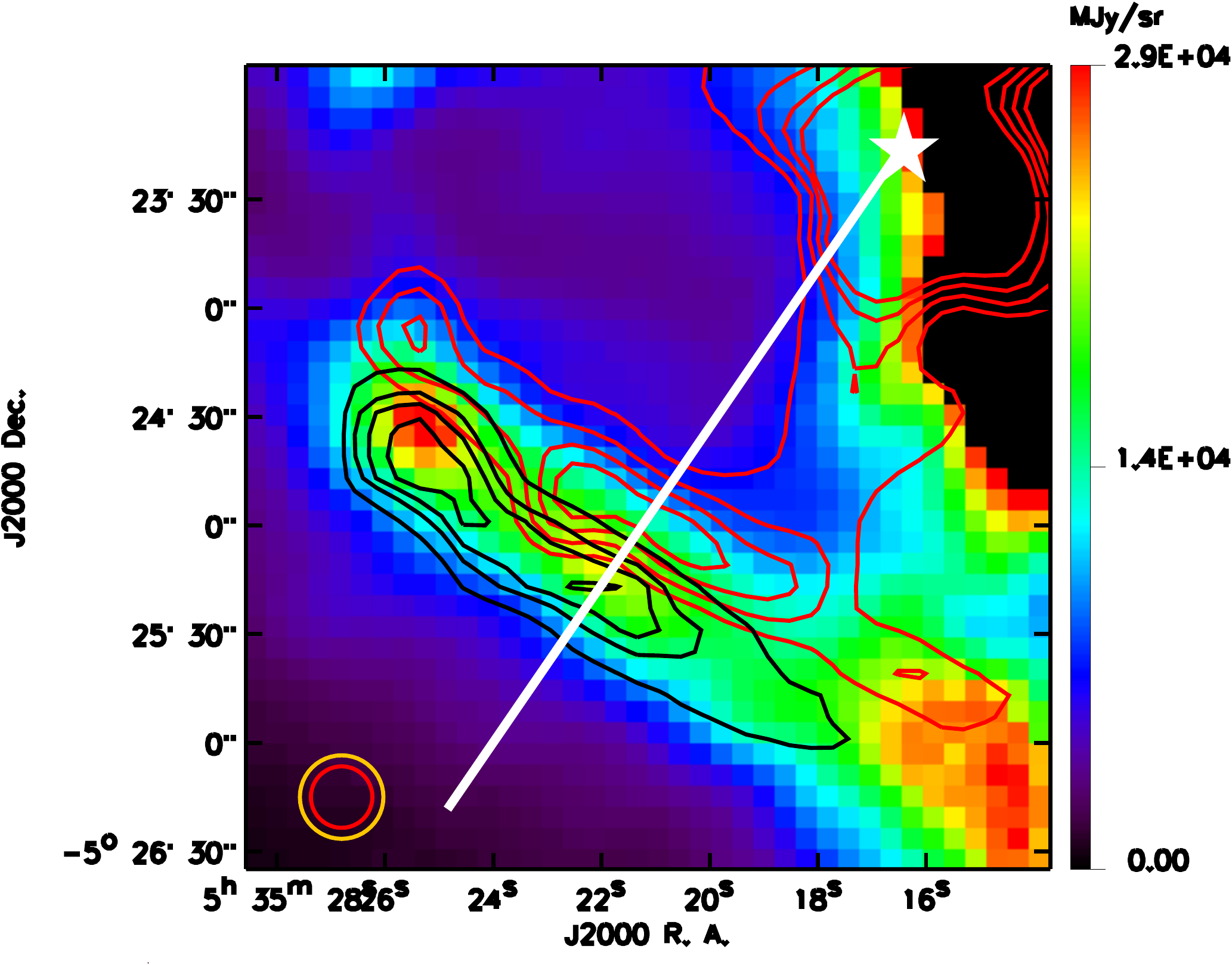}
    \vskip+0.2cm
      \caption{\emph{Herschel} 250\,$\mu$m observation (color image) and IRAC 3.6 $\mu$m map (red contours) in the SPIRE 250 $\mu$m beam (red circle; FWHM=17.6$\arcsec$). Black contours show the distribution of $\mathrm{C^{18}O}$ (8-7) observed with the SPIRE-FTS (yellow circle; FWHM=23.9$\arcsec$) from Habart et al. (in prep.). The white star illustrates the position of $\theta^1$ Ori C and the white line is the cut of Figure \ref{Fig:profile}.}
         \label{fig:3.6/250}
   \end{figure}

We extracted brightness profiles obtained along a cut going from the exciting star through the PDR, whose position is visible in Figure \ref{fig:3.6/250}. The Orion Bar appears as a prominent peak in these cross-cuts lying at $\sim 2\arcmin$ from $\theta^1$ Ori C (Fig. \ref{Fig:profile}). Sub-structure in this profiles appears at 3.6 and 70 $\mu$m showing that the Bar is more complicated than a simple ridge. A shift between the peak positions is also clearly visible. This shift is real as can be seen on the comparison of the 250\,$\mu$m map superimposed with contours of the 3.6\,$\mu$m map after convolution and reprojection as described above (Fig. \ref{fig:3.6/250}). The 250\,$\mu$m emission clearly rises further from the exciting star than the 3.6\,$\mu$m one. This can be generalized: the shorter the wavelength is, the closer to the exciting source is the emission. As we probe bigger grains at larger wavelengths, we see that the BG emission is located further the ionization front than the emission from the smaller grains. Moreover, we show in Figure \ref{fig:3.6/250} the C$^{18}$O emission observed by the SPIRE-FTS (Habart et al., in prep.), which traces the dense matter and a good correlation is observed with the 250\,$\mu$m broadband emission, confirming that the FIR dust emission is a tracer of the densest part of the PDR (see also \citealp{Buckle:2011}).\\
The ionization front is located at 111$\arcsec$ from $\theta^1$ Ori C (\cite{2009ApJ...693..285P} from [SII] intensity profile; \cite{2012A&A...538A..37B} from [NII] PACS observations). Adopting the same cut as \cite{2012A&A...538A..37B}, we find from the 3.6\,$\mu$m emission that PAHs peak at 116 $\pm 2$$\arcsec$ from the star, just behind the ionization front. We lack good VSG tracers since 24 $\mu$m \emph{Spitzer}/MIPS data of this region are unusable because of saturation and the 70 $\mu$m PACS emission is dominated by BGs. The peak of the BG emission is around 150\,$\arcsec$ far from the exciting star. The fact that the emission of different species rises successively when we enter the PDR is a result of the nearly edge-on orientation of the Orion Bar (see also \citealp{2009ApJ...705..226D}).   

 \begin{figure}[!ht]
 \centering
   \includegraphics[width=\columnwidth]{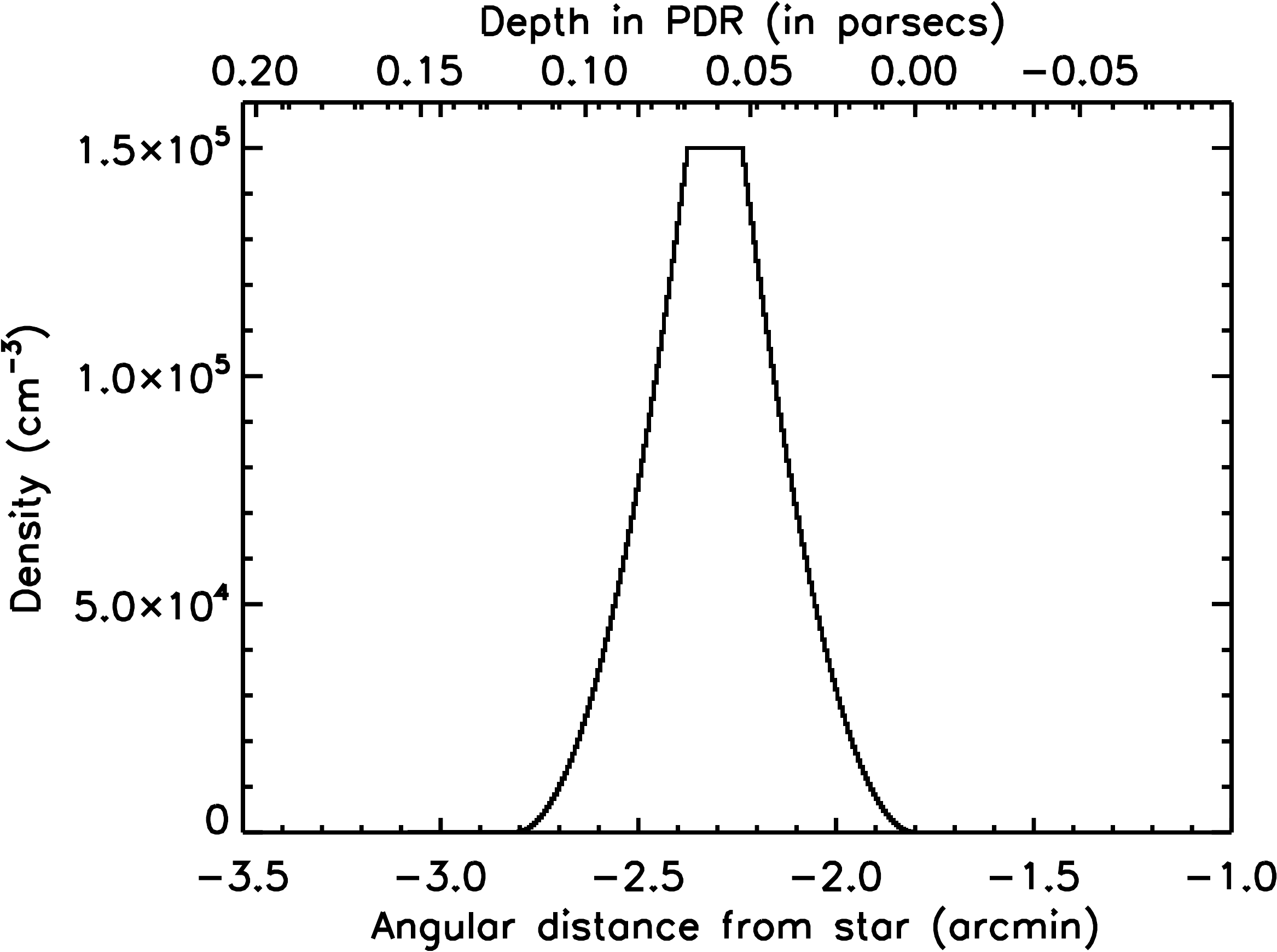}
      \caption{Density profile with a power-law rise and a symmetric fall. The maximum density is 1.5$\times 10^5$ cm$^{-3}$ and is reached at 5.25$\times10^{-2}$ pc far from the PDR edge.}
         \label{fig:density}
   \end{figure}

\section{Modelling}
\label{modelling}  

The Orion Bar is a very dense PDR which is illuminated by a strong radiation field, therefore the dust size distribution and optical properties should show some differences to those seen toward diffuse regions. Our goal is to extract evidence of this evolution, and to characterise and to quantify it. However, radiative transfer effects have a crucial influence on the dust emission, which is why modelling is needed to recover the properties of grains in the Orion Bar. 

\subsection{Model description}
We model the brightness profiles using the DustEM code for the dust emission coupled with a plane parallel radiative transfer code.\\
DustEM \citep{2011A&A...525A.103C} is a numerical tool which calculates the emission of an ensemble of dust populations of various properties under the effect of a given radiation field. The dust model we use includes four different populations: PAHs, small amorphous carbons (SamC), large amorphous carbons (LamC), and astronomical silicates (aSil). The dust abundances and emissivities used in the model are indicated in Tab.\,\ref{tab:model}. The reader is referred to Fig.\,1 of \cite{2011A&A...525A.103C} to get a detailed description of the size distribution and to Fig.\,4 and 5 in the same reference to examine the computed extinction curve.
In PDRs, evolution of the physical conditions is dominated by the changes in the radiation field. Therefore the radiative transfer effects are critical for our study and need to be properly quantified.  We have used the model described in \cite{2008A&A...491..797C} to take them into account. In this model, the PDR is represented by a semi-infinite plane-parallel slab whose density is defined by the user. The cloud is illuminated by the Mathis ISRF \citep{1983A&A...128..212M}, plus a blackbody radiation field derived from stellar parameters of $\theta^1$ Ori C (radius and distance to the PDR). Then the code computes the flux $F(z,\lambda$):
\begin{equation}
F(z,\lambda)=F_{t}(z,\lambda)+F_{b}(z,\lambda)+F_{IR}(z,\lambda),
\label{eq:model}
\end{equation}
where $z$ is the depth in the PDR, $\lambda$ is the wavelength, $F_{t}(z,\lambda)$ is the transmitted flux, $F_{b}(z,\lambda)$ is the backscattered one, and $F_{IR}(z,\lambda)$ denotes the emission coming from low depth dust, which absorbs, re-emits and thus heats the grains located deeper in the PDR. The calculation was performed assuming an asymmetry parameter for the scattering phase function $g$ of 0.6 which corresponds to the generally agreed value for the diffuse ISM \citep{1997ApJ...481..809W}. Moreover, the model only considers a single backscattering, in other words, when a photon is backscattered, it cannot go back toward the densest parts of the PDR. This approximation is justified as the $g$ value indicates a strong forward scattering (88$\%$ forward against 12\% back). Nevertheless, the main assumption of this model is the plane-parallel approximation. It greatly simplifies the computations since it allows a 1-D treatment.
 
The radiative code requires a density profile $n_{H}(z)$ for the PDR. For convenience, we use a power-law rise starting at $z_{0}$ and defined as: 
\begin{equation}
n_{H}(z)=n_{0}\times \left( \frac{z}{z_{0}} \right)^{\alpha},
\label{eq:dens}
\end{equation}
where $z_{0}$ is the depth at which the density reaches the maximum value $n_{0}$ and $\alpha$ is the power-law index. The density keeps constant at the maximum value between $z_{0}$ and $z_{1}$ and then decreases symmetrically. Indeed, models with a rise followed by a constant density do not reproduce the observed decrease of intensity after the peak at larger wavelengths. A lower density is required and for convenience we adopt a symmetric fall off. The density profile is showed in Figure \ref{fig:density}.\\
The four parameters, $n_{0}$, $\alpha$, $z_{0}$ and $z_{1}$ are free to vary in the fitting, contrary to the radiation field parameters (the radius of the exciting star, its temperature, and its distance from the cloud) which are fixed to give an illumination equal to $2.6\times10^{4} \ G_{0} $ in Mathis units, following \cite{1998A&A...330..696M}. 
Moreover, as the DustEM code computes the emission per hydrogen atom it is thus necessary to make an assumption on the column density to compare our model outputs to the data. It is made through the $l_{PDR}$ parameter, corresponding to the PDR length along the line of sight. It is related to the column density $N_{Htot}(z)$ by $N_{Htot}(z)=n_{H}(z)\ l_{PDR}$, where $n(l)$ is the density along the line of sight. This is a fifth parameter which is set to reproduce the emission given a density profile. $l_{PDR}$ has to be compatible with previous measurement of the column density in the Orion bar. $2\times10^{23}$ cm$^{-2}$ is an order of magnitude given by \cite{2003A&A...412..157J} from SCUBA/JCMT observations at 850 $\mu$m. \\
 \begin{figure*}
 \centering
   \includegraphics[width=15cm]{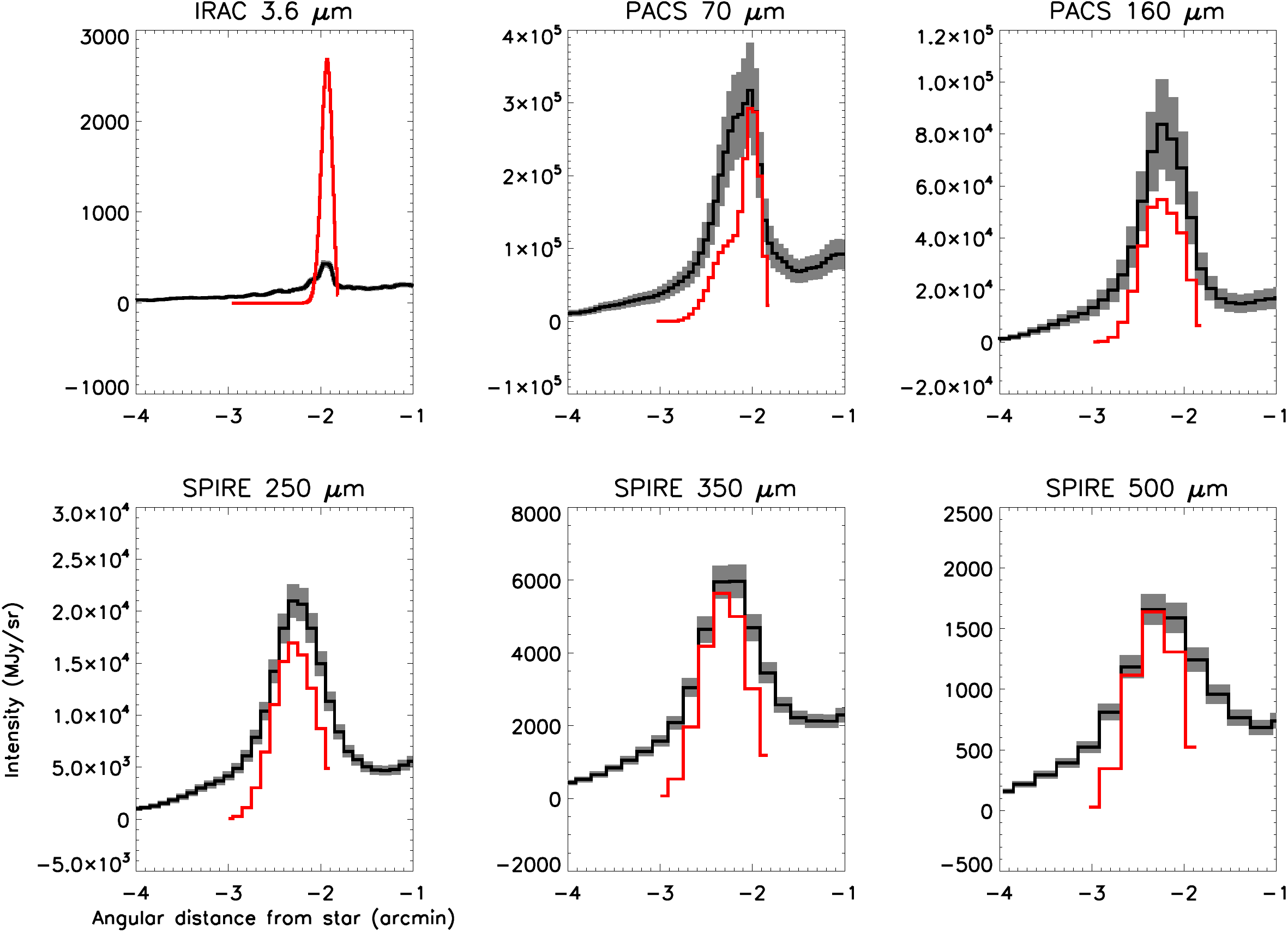}
      \caption{Brightness profiles from \emph{Spitzer} and \emph{Herschel} observations (black), The photometric uncertainties are respectively 5$\%$ for IRAC, 20$\%$ for PACS and 7$\%$ for SPIRE (grey). The modelling using diffuse ISM dust properties (Tab\,\ref{tab:model}) is shown by the red lines.}
         \label{profile_model1}
   \end{figure*}

 \begin{figure*}
 \centering
   \includegraphics[width=15cm]{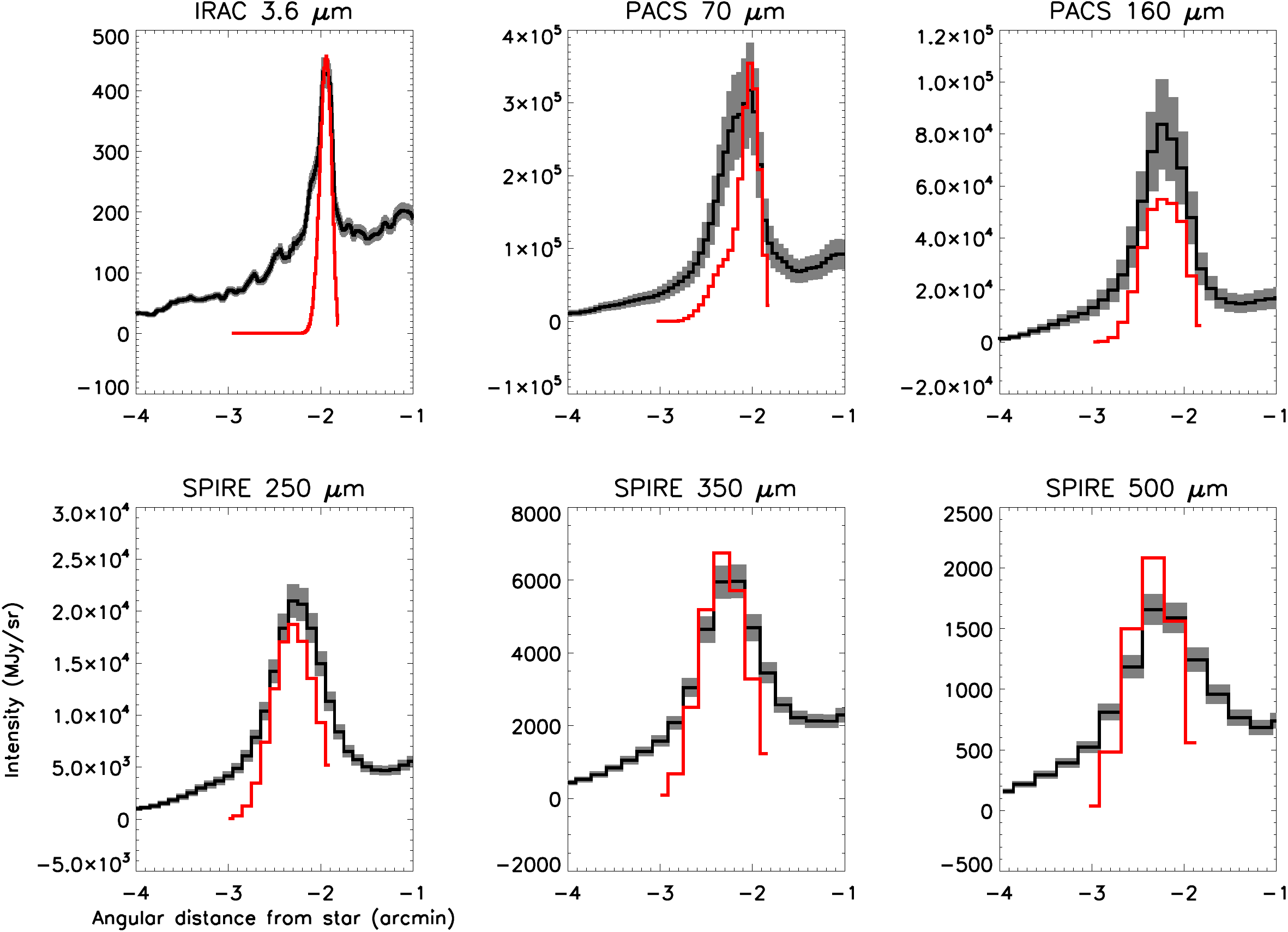}
      \caption{Same as Figure \ref{profile_model1} with a PAH abundance 7 times lower than in the diffuse ISM (Tab\,\ref{tab:model}). The increase of brightness in the model at long wavelengths is due to a weaker absorption in the UV/visible.}
         \label{profile_model2}
\end{figure*}

 \begin{figure*}
 \centering
   \includegraphics[width=15cm]{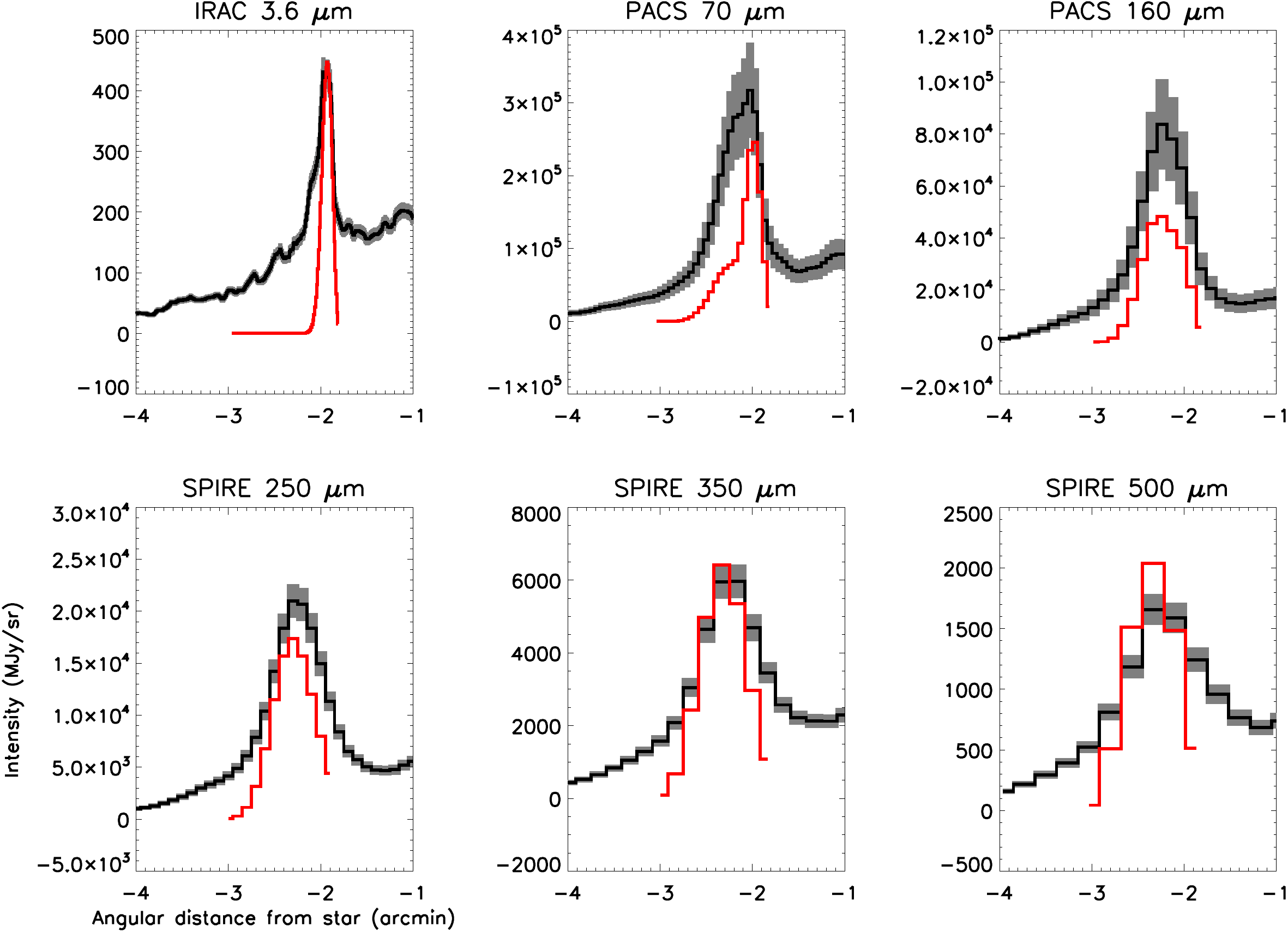}
      \caption{Same as Figure \ref{profile_model1} with $l_{PDR}=0.25\,\mathrm{pc}$. PAH abundance is a factor 3.3 lower than the canonical value in the diffuse ISM. The BG emissivity is enhanced by a factor 2 (Tab.\,\ref{tab:model}). }
         \label{profile_model3}
\end{figure*}
To take into account the self-absorption by dust, which can be important in the NIR, the emerging intensity is reduced by: 
\begin{equation}
\int_{0}^{l_{PDR}} \! \frac{e^{-\tau(\lambda,l)}}{l_{PDR}} dl,
\label{eq:selfabs}
\end{equation}
where $l$ is the distance along the line of sight ($l=0$ at the PDR edge) and $\tau(\lambda,l)$, the optical depth, computed by assuming no density gradient along the line of sight and using the extinction curve computed by DustEM. This factor is only significant at 3.6\,$\mu$m since the optical depth is actually very low at longer wavelengths ($\tau_{\mathrm{max}}<0.4$ at 70 $\mu$m). Finally, the model output is convoluted by the appropriate PSF and integrated over each considered filter to be compared to the data.
\subsection{Fitting}
Several different density profiles have been tested to explore the parameter space and the influence of each parameter on the dust emission profiles. We allow $n_{0}$ to vary between $10^4$ and $10^6$ cm$^{-3}$ which is the canonical density range for the Orion Bar (\citealp{1995A&A...294..792H, 1994ApJ...422..136T}). For each $n_{0}$ value, several power-law indexes $\alpha$ are considered between 1 (a linear rise of density) and 12 (a very steep profile). As for $z_{0}$ and $z_{1}$, they are well constrained by the maximum intensity position in each band and require to be fitted a posteriori given the values of $n_{0}$ and $ \alpha$.\\  
The maximum density $n_{0}$ affects the overall brightness. The densest is the PDR, the highest is the brightness in Herschel bands. For a given $n_{0}$ and at a given wavelength, the $\alpha$ power-law index defines where the dust emission peaks. Indeed, the brightness peak position depends on a competition between the incident flux and the density of absorbing particles.There is thus a degeneracy between $n_{0}$ and $\alpha$ and to keep a peak position constant with an increasing density, a higher power-law index is needed. We also notice that the couple $(n_{0},\alpha)$ changes the shape of the profiles. On the other hand, the length of the PDR along the line of sight is a multiplicative factor and affects each band in the same way.\,\,\,\,\,\,\,\,\,\, \\

Concerning the dust model, we consider the abundances and optical properties of dust in the diffuse ISM. The best adjustment of the data has been obtained with the following parameters: $n_{0}=1.5\times10^5 \ \mathrm{cm^{-3}}$, $\alpha=2$, $z_{0}=5.25\times10^{-2}\ \mathrm{pc}$, $z_{1}=6.95\times10^{-2} \ \mathrm{pc}$ and $l_{\mathrm{PDR}}=0.45 \ \mathrm{pc}$. Using these parameters we obtain $N_H=n_{0}\times\,l_{PDR}=2\times10^{23} \mathrm{cm^{-2}}$ which is of the same order of magnitude as measured by \cite{2003A&A...412..157J}. Putting the geometry all together leads to a ridge which is $\sim$0.45 pc long along the line of sight, $\sim$0.05 pc thick (width at half maximum of the density profile) and $\sim$0.42 pc long in the sky plane. 
Figure\,\ref{profile_model1} compares the observed and modelled brightness profiles. 
The spatial shift between the maximum intensity positions observed at different wavelengths is well reproduced. The absolute brightnesses in the \emph{Herschel} bands are also quite well reproduced by the model, except the 160 and 250 $\mu$m emissions which are underestimated. The modelled profiles are also narrower than the dat and, at shorter wavelengths, the PAH emission is significantly overestimated by the model. 

\renewcommand{\arraystretch}{1.5}
  \begin{table*}[ht]
      \caption[]{Properties of the different models tested.}
         \label{tab:model}
              	\vskip-0.5cm
     $$
         \begin{array}{|p{0.34\linewidth}|c|c|c|c|c|c|}
            \hline
          \  &\hspace{0.5cm}Y_{\mathrm{PAH}} \ (M/M_H) \hspace{0.5cm} &\hspace{0.5cm} Y_{\mathrm{SamC}} \hspace{0.5cm} & \hspace{0.5cm}Y_{\mathrm{LamC}} \hspace{0.5cm} &  \hspace{0.5cm}Y_{\mathrm{aSil}} \hspace{0.5cm}&\hspace{0.5cm} \epsilon_{\mathrm{FIR}}\hspace{0.5cm}  & \hspace{0.5cm}l_{\mathrm{PDR}} \mathrm{(pc)} \hspace{0.5cm}\\
            \hline
            Diffuse ISM model (Fig.\,\ref{profile_model1}) & 7.8\times10^{-4}   & 1.65\times10^{-4} & 1.45\times10^{-3} & 7.8\times10^{-3}& \epsilon^0_{\mathrm{FIR}}(\lambda)\ $\tablefootmark{a}$& 0.45 \\
             \hline
            PAH depleted model (Fig.\,\ref{profile_model2})     & 1.1\times10^{-4}   & 1.65\times10^{-4} & 1.45\times10^{-3} & 7.8\times10^{-3} & \epsilon^0_{\mathrm{FIR}}(\lambda)\ $\tablefootmark{a}$ & 0.45 \\
             \hline
         PAH depleted model + $\epsilon_{\mathrm{BG}}$ enhancement (Fig.\,\ref{profile_model3})    &  2.36\times10^{-4}   & 1.65\times10^{-4} & 1.45\times10^{-3} & 7.8\times10^{-3}& 2\times\epsilon^0_{\mathrm{FIR}}(\lambda)\ $\tablefootmark{a}$ & 0.25  \\
            \hline
            
         \end{array}
     $$ 
     \vspace{-0.5cm}
     \tablefoot{
     \tablefoottext{a}{$\epsilon^0_{\mathrm{FIR}}(\lambda)$ is the FIR emissivity presented in Fig.A1 from \cite{2011A&A...525A.103C}.}
     }
   \end{table*}


\section{Discussion}
\label{discussion}

Using our plane parallel model of the Orion Bar, assuming dust properties typical for the diffuse atomic ISM and with values of the local and column densities compatible with previous observations, we are able to reproduce very well the stratification of the dust emission inside the PDR seen in \emph{Herschel} and \emph{Spitzer}/IRAC data (Figs.\,\ref{fig:map} and \ref{Fig:profile}). This indicates that the increase of the distance between the observed Bar and the illuminating star with increasing wavelength is a pure radiative transfer effect. This model is also able to reproduce at first order the global shape of the brightness profiles and the absolute brightness observed with \emph{Herschel}. We conclude that a simple plane parallel model seems to be a reasonable approximation of the Bar.\\      
The discrepancy between the synthetic and observed brightness profiles in \emph{Herschel} bands can be explained by two assumptions we make. First, the PDR is modelled as a single slab without matter before and behind, and so the model does not reproduce the observed brightness outside the ridge. Second, the geometry may be not purely edge-on (e.g., \citealp{1993Sci...262...86T, 1995ApJ...438..784W, 2009ApJ...693..285P}). Inclination effects, which are not taken into account in our study, can broaden the brightness profiles and the SEDs. Sub-structures which are ignored in our modelling can also broaden the profiles.\\

\begin{figure}
 \centering
   \includegraphics[width=8.5cm]{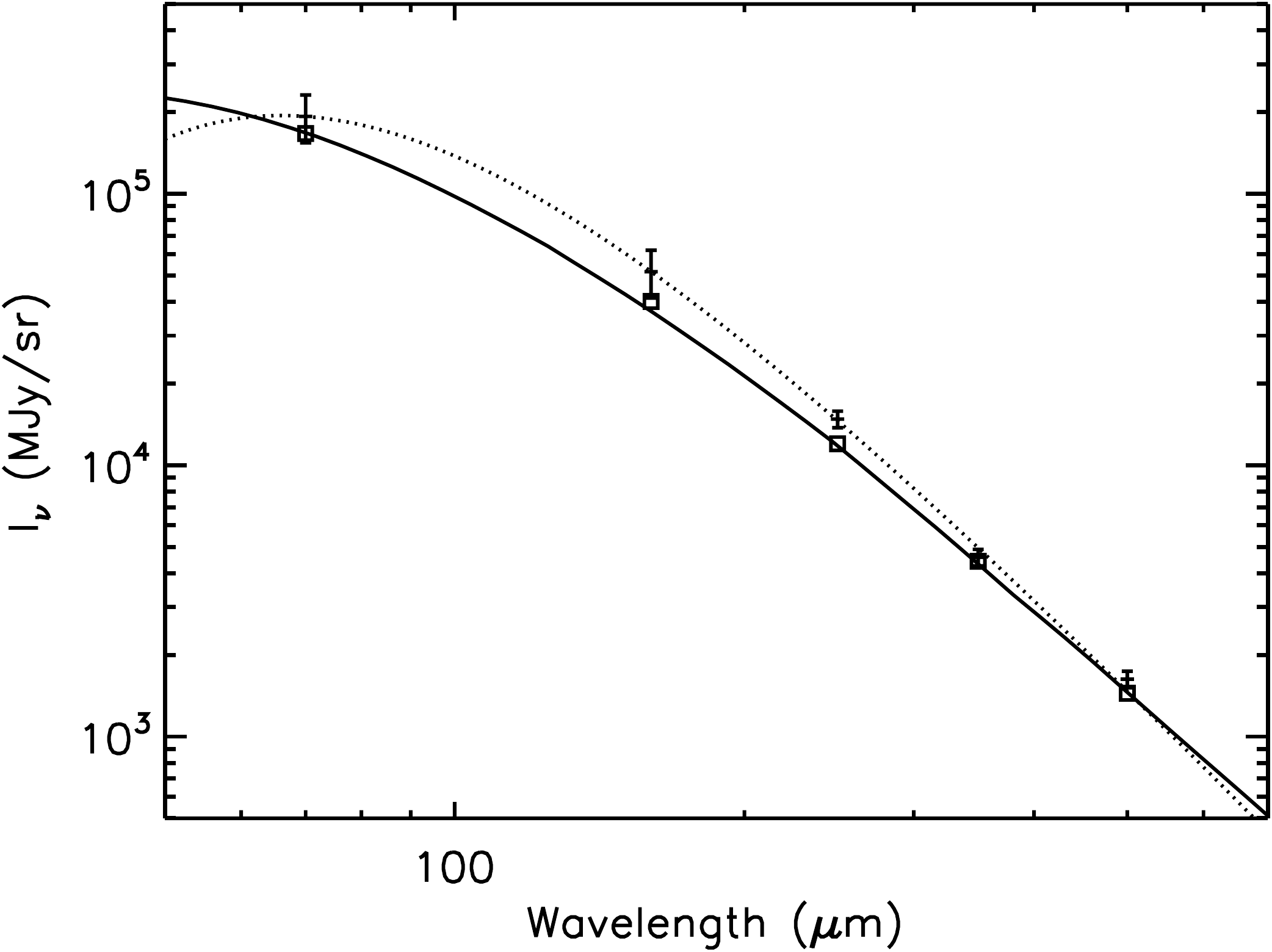}
      \caption{Solid line: SED extracted from the radiative transfer model using diffuse ISM properties (Tab.\,\ref{tab:model}) at the 70\,$\mu$m peak position (Fig\,\ref{profile_model1}). Square: Integrated fluxes in \emph{Herschel} bands. Dotted line: Modified blackbody fitted to the data.}
         \label{fig:fit_SED}
   \end{figure}

The overestimated PAH emission in the Orion Bar while using diffuse ISM dust abundances is not a surprise and could be due to several phenomena.
\begin{enumerate} 
\item Dust grains in the Orion bar undergo a very hard UV radiation field which can photo-destroy PAHs at the ionization front \citep{1994A&A...291..239G}. This could be coupled with PAH sticking on BGs in the molecular cloud \citep{1994ApJ...422..164K}. Reducing the PAH abundance by a factor 7, without changing any other parameter, allows a good match of the 3.6\,$\mu$m brightness profile (Fig.\,\ref{profile_model2}). Less PAHs reduces the UV absorption and therefore the emission in \emph{Herschel} bands slightly increases (maximum brightness multiplied by factors 1.1-1.2 except at 160\,$\mu$m where it is unchanged). 

\item In a dense medium, where the radiation does not penetrate, dust grains are suspected to be coagulated \citep{1989IAUS..135..239T}, which increases the overall FIR emissivity $\epsilon_{FIR}$ by factors about 2-3 (e.g., \citealp{1999A&A...347..640B, 2003A&A...398..551S, 2009A&A...506..745P, 2011A&A...528A..96K}). To be in agreement with the observations, a model with an increase of $\epsilon_{FIR}$ must be counterbalanced by a decrease of the column density (through the $l_{PDR}$ parameter) by the same factor. Taking into account this enhancement allows a good match of the brightness profiles, as illustrated in Figure \ref{profile_model3} and a decrease of the length along the line of sight by factor 2-3 is still compatible with observational constraints.
\item The length along the line of sight may be not constant. As mentioned above, in \emph{Herschel} bands we probe the densest part of the Bar whereas at 3.6 $\mu$m we are sensitive to the surface emission which is very structured (first panel of Fig.\,\ref{fig:map}). What appears as a bar in the sky plane could be a convex surface pointing to the illuminating star, and thus presenting a bigger depth along the line of sight in the densest part than near the ionization front. 

PDR models of the gas emission in the Orion Bar focus on the HII region and the zone near the ionization front, and fit the observations considering a length along the line of sight close to 0.15 pc \citep{2009ApJ...693..285P} while we take $l_{PDR}=0.45$ pc in our modelling. Therefore, the excess of the 3.6 $\mu$m emission in our model could be simply explained by a decrease of the length along the line of sight near the ionization front.
\end{enumerate}
Finally, the observed absolute brightnesses at 160\,$\mu$m and at 250\,$\mu$m are underestimated by factors 1.5 and 1.2, respectively. This  comes from the dust model itself. With DustEM, optical properties of the BG component, made of silicates and large amorphous carbons, create an emerging spectrum with a value of the spectral emissivity index $\beta$ close to 1.55. Cooling and radiative transfer cannot change this. As a consequence, we fail to reproduce the SED of BGs, whatever the position on the Bar; grains are systematically cooler and spectra are systematically steeper in the data than in the model, as illustrated at the 70\,$\mu$m peak position shown in Figure \ref{fig:fit_SED}. SEDs of BGs at different position across the Bar (studied Sect.\,\ref{subsect:BGspec} and shown Fig.\,\ref{Fig:bgspec}) reveal $\beta$ values around 1.8 in the Bar. This discrepancy between the model and the data induces the underestimate of $\beta$ seen in Figure\,\ref{fig:fit_SED}.\\     

\section{Conclusion}

We have presented the first observations of the Orion Bar photodissociation region using the \emph{Herschel Space Observatory} in five broad bands (70 and 160 $\mu$m with PACS, 250, 350 and 500 $\mu$m with SPIRE). 
The wavelength range and the spatial resolution of \emph{Herschel} allow us to map and resolve the emission of the biggest grains located in the densest zones of the Orion Bar. \\
To extract SEDs, we have developed an original and efficient method to compute convolution kernels allowing to convert all maps to the same beam. Study of BG SEDs at different positions in the Orion Bar region reveals a cooling of dust (from 80 K to 40 K) and an increase of the spectral emissivity index $\beta$ (from 1.1 to 2) as we go deep into the PDR. Combining the new \emph{Herschel} observations with ancillary data from \emph{Spitzer}/IRAC gives us the opportunity to probe the morphology of the Orion Bar from 3.6\,$\mu$m to 500\,$\mu$m. The global structure is the same at all wavelengths but we observe an increase in the distance star-Bar with increasing wavelength (shift of $\sim$ 30" between the 3.6\,$\mu$m and the 500\,$\mu$m profiles), which implies a stratification of dust properties within the Bar. 

We have used a radiative transfer model using typical diffuse ISM dust properties and abundance to model the brightness profiles observed with \emph{Herschel} and \emph{Spitzer}. This model reproduces the stratification within the Bar using values of the column density and local density parameters that are in agreement with the observational constraints. This shows that this stratification is purely a radiative transfer effect.

However, using diffuse ISM dust properties and abundances in the model leads to several discrepancies in shape and absolute brightness between the modelled brightness profiles and the observations. The assumptions we make about the geometry (considering the PDR as a single slab, adoption of an edge-on geometry, without inclination and sub-structure effects) can explain the differences of width between the synthetic and observed profiles. \\
The relatively low absolute brightness at 3.6\,$\mu$m might be explained by PAH photo-destruction by the strong UV radiation, or by sticking of PAHs to the BGs in the pre-existing molecular cloud. A PAH abundance decrease by a factor 7 from the canonical value in the diffuse ISM allows a good fit of the IRAC 3.6\,$\mu$m emission.  
A combination of a decreased PAH abundance (by a factor less than 7) with a decreased length along the line of sight is possible too. If so, an increase of the brightness in \emph{Herschel} bands is needed to fit the data. Coagulation can explain such a behaviour since it can enhance the FIR emissivity by factors 2-3.  
An alternative is that the length along the line of sight is shorter at 3.6\,$\mu$m since we probe the emission from the surface of dense structures.


\begin{acknowledgements}
We are grateful to the anonymous referee for his or her careful reading and for the suggestions which helped to improve the quality of the manuscript.  We thank Laurent Verstraete and Mathieu Compi\`egne for useful discussions about the DustEM model. HIPE is a joint development by the Herschel Science Ground Segment Consortium, consisting of ESA, the NASA Herschel Science Center, and the HIFI, PACS and SPIRE consortia. PACS has been developed by a consortium of institutes led by MPE (Germany) and including UVIE (Austria); KU Leuven, CSL, IMEC (Belgium); CEA, LAM (France); MPIA (Germany); INAF-IFSI/OAA/OAP/OAT, LENS, SISSA (Italy); IAC (Spain). This development has been supported by the funding agencies BMVIT (Austria), ESA-PRODEX (Belgium), CEA/CNES (France), DLR (Germany), ASI/INAF (Italy), and CICYT/MCYT (Spain). SPIRE has been developed by a consortium of institutes led by Cardiff University (UK) and including Univ. Lethbridge (Canada); NAOC (China); CEA, LAM (France); IFSI, Univ. Padua (Italy); IAC (Spain); Stockholm Observatory (Sweden);   Imperial College London, RAL, UCL-MSSL, UKATC, Univ. Sussex, STFC, UKSA (UK); and Caltech, JPL, NHSC, Univ. Colorado (USA). This development has been supported by national funding agencies: CSA (Canada); NAOC (China); CEA, CNES, CNRS (France); ASI (Italy); MCINN (Spain); SNSB (Sweden); STFC (UK); and NASA (USA).
    \end{acknowledgements}

\bibliographystyle{aa}
\bibliography{biblio_oribar_observations}

\end{document}